\documentclass[aps,preprint,prb,showpacs,tightenlines]{revtex4}

\usepackage{epsfig}

\newcommand{\smfrac}[2]{\mbox{$\frac{#1}{#2}$}}

\begin{document}

\title{Unified description of ballistic and diffusive carrier transport
\\ in semiconductor structures}

\author{R.\ Lipperheide}
\author{U.\ Wille}
\email{wille@hmi.de}

\affiliation{Abteilung Theoretische Physik, Hahn-Meitner-Institut
Berlin,\\ Glienicker Str.\ 100, D-14109 Berlin, Germany}

\date{\today}

\begin{abstract}
A unified theoretical description of ballistic and diffusive carrier transport
in parallel-plane semiconductor structures is developed within the
semiclassical model.  The approach is based on the introduction of a
thermo-ballistic current consisting of carriers which move ballistically in the
electric field provided by the band edge potential, and are thermalized at
certain randomly distributed equilibration points by coupling to the background
of impurity atoms and carriers in equilibrium.  The sum of the thermo-ballistic
and background currents is conserved, and is identified with the physical
current.  The current-voltage characteristic for nondegenerate systems and the
zero-bias conductance for degenerate systems are expressed in terms of a
reduced resistance.  For arbitrary mean free path and arbitrary shape of the
band edge potential profile, this quantity is determined from the solution of
an integral equation, which also provides the quasi-Fermi level and the
thermo-ballistic current.  To illustrate the formalism, a number of simple
examples are considered explicitly.  The present work is compared with previous
attempts towards a unified description of ballistic and diffusive transport.

\end{abstract}


\maketitle

\section{Introduction}

Within the semiclassical model, charge carrier transport in semiconductor
structures is commonly described in two limiting situations.  In the ballistic
limit (carrier mean free path much larger than the characteristic dimensions of
the sample), the carriers move unscattered all through the sample.  The
conduction is treated as transmission across the band edge profile of carriers
emitted and absorbed at external contacts which are represented by large
reservoirs in thermal equilibrium \cite{bet42,sha65,lan87,dat95,imr99}.  This
is the picture of ballistic motion thermally activated at the contacts in terms
of which one derives a current-voltage characteristic for the current flowing
in the sample.  In the other, the diffusive, limit (mean free path much smaller
than the characteristic dimensions of the sample), one has the Drude model of
carriers moving from one equilibrium state to another, infinitesimally
close-lying one, each characterized by the local quasi-Fermi level and by the
electric field provided by the band edge potential
\cite{max91,dru00,sch38,mot38,ash76,sap95}.  This yields the local
drift-diffusion equations for the current density in the sample.

In the present work, we consider the general case where the ratio between mean
free path and the characteristic dimensions of the sample may assume any given
value.  For definiteness, we treat electron transport; the case of holes is
analogous.  We develop a unified description of ballistic and diffusive
electron transport in parallel-plane semiconductor structures, in which the
ballistic motion is thermally activated neither only at the distant external
contacts nor only at infinitesimally close-lying points, but at separations
anywhere between these two limits.  No restrictions are imposed on the shape of
the conduction band edge profile.

The unified description relies on the introduction of a ``thermo-ballistic
current'', where the electrons move ballistically in the band edge potential
over the length of certain randomly distributed ``ballistic intervals'', at the
ends of which they are thermalized.  The probability for ballistic
(unscattered) motion between any two end points (`` points of equilibration'')
is governed by the mean free path.  The thermalization is effected by the
coupling of the electron motion to the background of impurity atoms and
electrons in equilibrium.

The equilibration at the end points of the ballistic intervals, i.e., at the
points separating adjacent intervals, is viewed as absorption of the electrons
impinging from either side, and re-emission with a momentum distribution
determined by the local quasi-Fermi level of the background.  The absorbed and
re-emitted currents do not, in general, counter-balance, and produce a
background current whose local increment is equal to the negative of the
increment of the thermo-ballistic current.  Therefore, the sum of the
thermo-ballistic and background currents is conserved, and is to be identified
with the physical current.  From the condition that the integral of the
background current over the length of the sample vanishes, an integral equation
is derived for the quasi-Fermi level inside the sample in terms of a
``resistance function'' $\chi$, which yields the thermo-ballistic current and
whose value $\widetilde{\chi}$ at one end of the sample determines the
current-voltage characteristic.  This quantity depends on the mean free path
$l$ and on the shape of the conduction band edge profile $E_{\rm c}$,
$\widetilde{\chi} = \widetilde{\chi}(l,[E_{\rm c}])$, and is called the
``reduced resistance''. It is the central quantity of the unified description
of ballistic and diffusive electron transport.  The derivation is carried out
for the nondegenerate case; it turns out that essentially the same reduced
resistance determines the zero-bias conductance in the degenerate case.

In earlier papers \cite{cro66,sze81,eva91,dej94,pri98}, attempts have been made
to treat the synthesis of ballistic and diffusive transport for particular band
edge profiles.  A heuristic unified model yielding a closed expression for the
reduced resistance in terms of the shape of an {\em arbitrary} band edge
profile has been presented in Ref.\ \cite{lip01}.  The present work is directed
towards a comprehensive description of electron transport, with emphasis on a
consistent theoretical formulation.  No attempt is made here to apply our
approach in the analysis of experimental data.

In the next section, we introduce the two basic elements of our description,
the points of local equilibration and the ballistic intervals, and derive the
probabilities of occurrence of the latter.  In Sec.\ III, we construct the
thermo-ballistic current and establish its connection with the physical
electron current.  In Sec.\ IV, the integral equation for the function $\chi$
is derived and the current-voltage characteristic is expressed in terms of the
reduced resistance $\widetilde{\chi}$.  The ballistic and diffusive limits are
obtained as special cases.  A discussion of the degenerate case completes the
section.  Simple examples illustrating our formalism are considered in Sec.\ V.
We examine various types of conduction band edge profile:\ flat profile
(constant electrostatic potential), profiles rising (or falling) monotonically,
and profiles containing a single maximum.  We present calculations of the
reduced resistance $\widetilde{\chi}$, of the quasi-Fermi level, and of the
thermo-ballistic current for one-dimensional and three-dimensional transport.
In Sec.\ VI, we compare our approach with previous attempts to unify ballistic
and diffusive transport.  Finally, Sec.\ VII contains a summary of the present
work and an outlook.

\section{Equilibration points and ballistic intervals}

\begin{figure}
\epsfysize=12.0cm
\epsfbox[20 300 501 700]{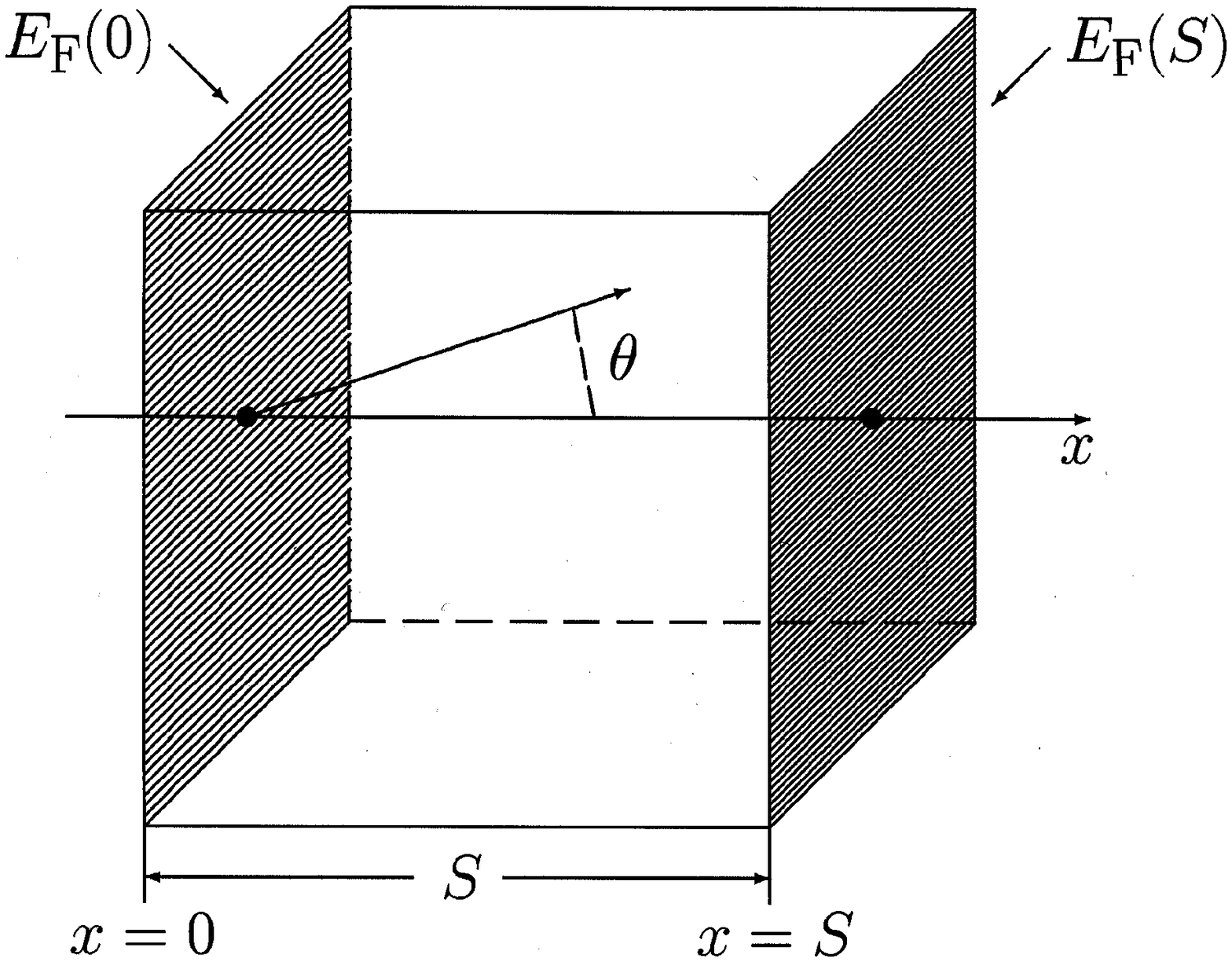}
\vspace*{4.0cm}
\caption{Schematic diagram showing a sample of length $S$ enclosed between
two plane-parallel contacts with associated Fermi levels $E_{\rm F}(0)$ and
$E_{\rm F}(S)$.}
\label{fig:1}
\end{figure}

We consider the transport of electrons in a semiconductor sample of the form of
a rectangular slab of thickness $S$, confined between two plane-parallel
contacts (cf.\ Fig.\ \ref{fig:1}).  The $x$-axis is taken normal to the planes
of the contacts, and the sample is assumed to be uniform in all directions
perpendicular to the $x$-axis.  We assume that the scattering at the boundary
planes parallel to the $x$-axis is specular so that effectively these planes
can be regarded as infinitely removed.  We are, therefore, dealing with a
one-dimensional geometry along the $x$-axis but, of course, the electrons move
in three-dimensional space.  For a two-dimensional electron gas, the latter may
be reduced to two dimensions, and for illustrative purposes one may even
consider electrons in one-dimensional motion (in or against the $x$-direction).

The contacts at $x=0$ and $x=S$ are viewed \cite{lan87,dat95,imr99} as large
reservoirs with Fermi energies $E_{\rm F}(0)$ and $E_{\rm F}(S)$, respectively,
whose difference determines the voltage bias $V$.  As a relaxation mechanism
inside the sample, we consider only isotropic impurity scattering characterized
by a universal mean free path $l$, which depends on the doping concentration,
but is independent of the electron velocity.  In the present geometry, the
spatial dependence of all quantities of interest reduces to that on the
$x$-coordinate.  For notational convenience, we introduce the scaled coordinate
$\xi = x/l$, that is {\em we measure all lengths in multiples of the mean free
path $l$}; the scaled coordinate $\sigma = S/l$ denotes the position of the
contact at $x=S$.

We introduce {\em points of equilibration} along the $\xi$-axis with the
understanding that these are actually short intervals (with position $\xi$ and
length $d\xi$) representing the three-dimensional sheet enclosed between two
(infinitely extended) planes perpendicular to the $\xi$-axis and separated by
the distance $d\xi$.  The electrons arriving at such points escape into the
background and are instantaneously replaced with background electrons emitted
isotropically into the thermo-ballistic current.  The density $J_{\rm e}(\xi)$
of the {\em particle current} emitted at a point $\xi$ [for the charge current
density $j_{\rm e}(\xi)$, one has to multiply by $-e$] into the right (left)
direction is given by
\begin{equation}
J_{\rm e}(\xi) = \pm \, v_{\rm e} \, n(\xi) \; ,
\label{eq:1}
\end{equation}
where $v_{\rm e} = (2 \pi m^{*} \beta)^{-1/2}$ is the emission velocity, and
$n(\xi)$ is the electron density at the equilibration point $\xi$.  For
a nondegenerate system (we consider the degenerate case later on), we have
\begin{equation}
n(\xi) = N_{\rm c} \, e^{-\beta [E_{\rm c}(\xi)- E_{\rm F}(\xi)]} \; ,
\label{eq:2}
\end{equation}
where $N_{\rm c} = 2(2\pi m^* / \beta h^2)^{3/2}$ is the effective density of
states at the conduction band edge, $E_{\rm c}(\xi)$ is the profile of the
conduction band edge potential, $E_{\rm F}(\xi)$ is the quasi-Fermi level
at the position $\xi$ in the interval $0 < \xi < \sigma$, and $\beta = 1/k_{\rm
B}T$.  The contacts at $\xi = 0$ and $\xi = \sigma$ may be regarded as special
points of equilibration characterized by {\em true} Fermi energies $E_{\rm
F}(0)$ and $E_{\rm F}(\sigma)$.  Including these quantities in the definition
of $E_{\rm F}(\xi)$, we continue to call this function the quasi-Fermi level
(now defined for $0 \leq \xi \leq \sigma$).

Between two points $\xi'$ and $\xi$ of local equilibration, the electrons move
ballistically in the electric field provided by the conduction band edge.
These {\em ballistic intervals} have an average length of the order of the mean
free path.  In order to determine the probability of occurrence of a ballistic
interval of length $|\xi- \xi'|$, we briefly digress and introduce the
following probabilities, which depend on the ratio of the distance between the
two equilibration points and the mean free path, i.e., on the scaled distance
$|\xi- \xi'|$:\

\noindent
1) --- the probability  $p_{1}(|\xi- \xi'|)$ that an electron emitted
at the point of local equilibration $\xi'$ reaches the point $\xi$ without
scattering [clearly $p_{1}(0) = 1$];

\noindent
2) --- the probability $p_{2}(|\xi- \xi'|)d\xi'$ that an electron emitted at a
point $\xi'$ after having been equilibrated there within the interval $d\xi'$
reaches the point $\xi$ without scattering and, analogously, the probability
$p_{2}(|\xi- \xi'|)d\xi$ that an electron emitted at $\xi'$ reaches the point
$\xi$ without scattering, and is equilibrated there within the interval $d\xi$;

\noindent
3) --- the probability $p_{3}(|\xi- \xi'|)d\xi'd\xi$ that an electron emitted
at a point $\xi'$ after having been equilibrated there within the interval
$d\xi'$ reaches the point $\xi$ without scattering, and is equilibrated there
within the interval $d\xi$.

The probability functions $p_{n}(\xi)$ depend on the dimensionality of the
electron motion via the averaged projection of the mean free path $l$ on the
$x$-axis (cf.\ below).

The probability $p_{2}(|\xi-\xi'|)d\xi$ is the product of the probability
$p_{1}(|\xi-\xi'|)$ for motion between $\xi'$ and $\xi$ without scattering
times the absolute value of the relative change of this "no-scattering
probability" (which is the negative of the relative change of the scattering
probability) within the interval $[\xi,\xi+d\xi]$,
\begin{equation}
p_{2}(|\xi-\xi'|)d\xi = p_{1}(|\xi-\xi'|)
\left(-\frac{dp_{1}(|\xi-\xi'|)}{p_{1}(|\xi- \xi'|)}\right) = -
dp_{1}(|\xi-\xi'|) \; , \label{eq:3}
\end{equation}
or
\begin{equation}
p_{2}(\xi)  =  - dp_{1}(\xi)/d\xi \; ; \ \ \  \xi \geq 0 \; .
\label{eq:4}
\end{equation}
Similarly, we have
\begin{equation}
p_{3}(|\xi-\xi'|)d\xi'd\xi  = - dp_{2}(|\xi-\xi'|) d\xi' \; ,
\label{eq:5}
\end{equation}
or
\begin{equation}
p_{3}(\xi)  =  - dp_{2}(\xi)/d\xi \; ; \ \ \  \xi \geq 0 \; .
\label{eq:6}
\end{equation}
For later use, we define more generally
\begin{equation}
p_{n}(\xi)  =  - dp_{n-1}(\xi)/d\xi \; ; \ \ \  \xi \geq 0 \; , \ \ \ n= 1,2,3
\; , \label{eq:7}
\end{equation}
or
\begin{equation}
\int_{0}^{\xi} d\xi' p_{n}(\xi') = p_{n-1}(0) - p_{n-1}(\xi) \; ; \ \ \ n=
1,2,3 \; ; \label{eq:8}
\end{equation}
here, the so far undefined quantity $p_{0}(0)$ is found, with the help of Eq.\
(7), to be given by
\begin{equation}
p_{0}(0) = \int_{0}^{\infty} d\xi \xi p_{2}(\xi) = \frac{1}{l}
\int_{0}^{\infty} \frac{dx}{l} x p_{2}(x/l) = \frac{\langle l \rangle}{l} \; .
\label{eq:9}
\end{equation}
The quantity $\langle l \rangle$ is the effective mean free path along the
$x$-axis; it depends on the dimensionality of the electron motion via the
probability function $p_{2}(x/l)$.

We now return to the probability of occurrence of the ballistic intervals.  The
latter have been defined as intervals of ballistic motion with equilibration
(absorption and emission) at the two ends.  Three different kinds of ballistic
interval are to be distinguished:\

\noindent
1) In the case of the ballistic interval spanned between the contacts at
$\xi = 0$ and $\xi = \sigma$, the electrons are emitted and absorbed,
respectively, at the contacts.  The probability of occurrence of this interval
is given by $p_{1}(\sigma)$.

\noindent
2) For the ballistic intervals $[0,\xi]$ and $[\xi,\sigma]$, the
equilibration occurs in the infinitesimal interval $[\xi,\xi + d\xi]$, and at
the contacts at $\xi = 0$ and $\xi = \sigma$, respectively.  These ballistic
intervals occur with probability $p_{2}(\xi) d\xi$ and $p_{2}(\sigma - \xi)
d\xi$, respectively.

\noindent
3) Finally, for the ballistic interval $[\xi,\xi']$, the probability of
occurrence is given by \linebreak $p_{3}(|\xi - \xi'|) d\xi' d\xi$.

Considering the explicit form of the probability functions  $p_{n}(\xi)$ for
the different dimensionalities, we find for {\em one-dimensional} motion
\begin{equation}
p_{n}(\xi) =  e^{-\xi} \; ; \ \ \  n=0,1,2,3 \; .
\label{eq:57}
\end{equation}
For $n=1$, this is the usual ``survival probability'' for ballistic motion
\cite{ash76,sap95}; the other cases follow with the help of Eqs.\
(\ref{eq:7})-(\ref{eq:9}).  In particular, $p_{0}(0) = 1$, which confirms
the obvious result $\langle l \rangle = l$.

For {\em two-dimensional} motion, we have to replace the mean free path $l$
with its projection $l \cos \theta$ on the $x$-axis (cf.\ Fig.\ \ref{fig:1}).
Then, changing on the right-hand side of Eq.\ (\ref{eq:57}) for $n=1$ the
variable $\xi$ to $\xi/\cos \theta$ and taking the average over the angle
$\theta$, we obtain
\begin{equation}
p_{1}(\xi) = \int_{0}^{\pi/2} d \theta \cos \theta \,  e^{-\xi/ \cos
\theta} = \int_{1}^{\infty} dt \, \frac{e^{-\xi t}}{t^{2} \sqrt{t^{2} - 1}} \;
, \label{eq:64}
\end{equation}
where we have introduced the variable $t = 1/\cos \theta$. More generally, we
have from Eqs.~(\ref{eq:7})-(\ref{eq:9})
\begin{equation}
p_{n}(\xi) = \int_{1}^{\infty} dt  \frac{e^{-\xi t}}{t^{3-n} \sqrt{t^{2} - 1}}
\; ; \ \ \ n=0,1,2,3 \; .
\label{eq:65}
\end{equation}
In particular,
\begin{equation}
p_{3}(\xi) =  K_{0}(\xi) \; ,
\label{eq:66}
\end{equation}
where $ K_{0}(\xi)$ is a modified Bessel function \cite{abr65}. Furthermore,
\begin{equation}
p_{0}(0) = \frac{\pi}{4} \; , \ \ p_{1}(0) = 1 \; , \ \ p_{2}(0) =
\frac{\pi}{2} \; ,
\label{eq:67}
\end{equation}
and the effective mean free path comes out as $\langle l \rangle = (\pi/4)
\, l$.

In the case of {\em three-dimensional} motion, we have
\begin{eqnarray}
p_{1}(\xi) &=& 2 \int_{0}^{\pi/2}d \theta \sin \theta \cos \theta \, e^{-\xi/
\cos \theta} = 2 \int_{0}^{1}d(\cos \theta) \, \cos \theta \, e^{-\xi/\cos
\theta} \nonumber \\ &=& 2 \int_{1}^{\infty} dt \, \frac{e^{-\xi t}}{t^{3}} = 2
E_3(\xi) \; .
\label{eq:68}
\end{eqnarray}
Here, $E_{n}(\xi) $ is the exponential integral of order $n$;\cite{abr65} the
factor 2 is a normalization factor, since
\begin{equation}
\int_{0}^{1}d(\cos \theta)  \cos \theta =   \frac{1}{2} \; .
\label{eq:69}
\end{equation}
Using Eqs.\ (\ref{eq:7})-(\ref{eq:9}) and the properties of the exponential
integral, we find
\begin{equation}
p_{n}(\xi) = 2 E_{4-n}(\xi) \; ;  \ \ \ n = 0,1,2,3 \; ,
\label{eq:70}
\end{equation}
and
\begin{equation}
p_{n}(0) = \frac{2}{3-n} \; ; \ \ \ n = 0,1,2 \; .
\label{eq:71}
\end{equation}
The effective mean free path is $\langle l \rangle = (2/3) \, l$.

\section{Thermo-ballistic current and physical current}

\begin{figure}
\epsfysize=12.0cm
\epsfbox[20 300 501 700]{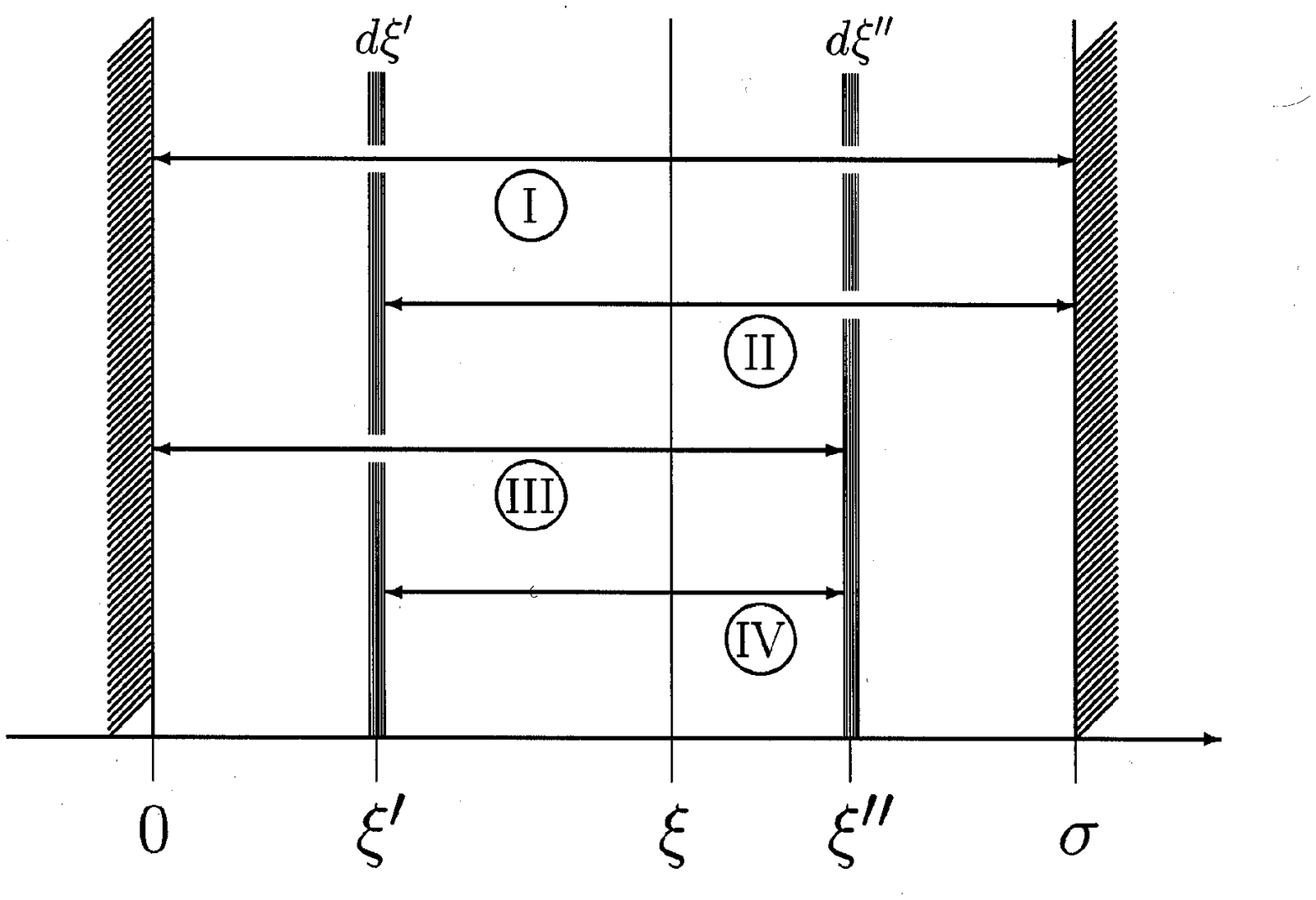}
\vspace*{4.0cm}
\caption{Diagram illustrating the four types of ballistic current entering into
the right-hand side of Eq.\ (\ref{eq:14}).}
\label{fig:2}
\end{figure}

The concepts established in the preceding section provide the basis for the
derivation of the net current at an arbitrary point $\xi$ $(0 < \xi < \sigma)$
inside the sample of length $\sigma$ (cf.\ Fig.~\ref{fig:2}).  On the one hand,
electrons arrive ballistically from the left, emitted at the left contact at
$\xi=0$ or at a point of local equilibration $\xi' \ (0 < \xi' < \xi)$; after
having passed the point $\xi$, they travel ballistically to the right contact
of the sample at $\xi = \sigma$ or to points of local equilibration at $\xi'' \
(\xi < \xi'' < \sigma)$.  The corresponding currents are illustrated by the
four arrows with heads pointing to the right.  On the other hand, electrons
arrive ballistically from the right, emitted at the contact at $\xi=\sigma$ or
at a point of local equilibration $\xi'' \ (\xi < \xi'' < \sigma) $, after
which they move on to the left contact at $\xi = 0$, or to points of local
equilibration at $\xi' \ (0 < \xi' < \xi)$, as illustrated by the arrows with
heads pointing to the left.

Inside some ballistic interval $[\xi',\xi'']$, the current emitted at $\xi'$
is determined by the electron density $n(\xi')$, and the current emitted at
$\xi''$, by $n(\xi'')$ [cf.\ Eq.\ (\ref{eq:1})].  However, owing to the
presence of the band edge potential $E_{\rm c}(\xi)$, part of the current is
reflected, so that we must introduce transmission probabilities
\begin{equation}
T_{\xi'}(\xi',\xi'')= e^{-\beta[E_{\rm c}^{\rm m}(\xi',\xi'') - E_{\rm
c}(\xi')]} \; , \; \; T_{\xi''}(\xi',\xi'')= e^{-\beta[E_{\rm c}^{\rm
m}(\xi',\xi'') - E_{\rm c}(\xi'')]}
\label{eq:10}
\end{equation}
for emission at the end points $\xi'$ and $\xi''$, respectively, of the
ballistic interval; here, $E_{\rm c}^{\rm m}(\xi',\xi'')$ is the maximum of the
potential profile within $[\xi',\xi'']$ (note that $E_{\rm c}^{\rm
m}(\xi',\xi'')$ is symmetric with respect to the interchange of $\xi'$ and
$\xi''$).  The transmission probabilities (\ref{eq:10}) are the classical ones;
if tunneling effects are important, the former should be replaced with the
corresponding WKB expressions (cf.\ Refs.\ \cite{dol54,str62}).  In view of
Eqs.\ (\ref{eq:1}) and (\ref{eq:2}), we now have $(0 \leq \xi' < \xi'' \leq
\sigma)$
\begin{equation}
J_{\rm l}(\xi', \xi'') = v_{\rm e} n(\xi') T_{\xi'}(\xi',\xi'') = v_{\rm
e} N_{\rm c} e^{-\beta [E_{\rm c}^{\rm m}(\xi',\xi'')-E_{\rm F}(\xi')]}
\label{eq:11}
\end{equation}
for that part of the current across $[\xi',\xi'']$ which is emitted on the left
at $\xi'$, and
\begin{equation}
J_{\rm r}(\xi', \xi'')= - v_{\rm e} n(\xi'') T_{\xi''}(\xi',\xi'') = - v_{\rm
e} N_{\rm c} e^{-\beta [E_{\rm c}^{\rm m}(\xi',\xi'')-E_{\rm F}(\xi'')]}
\label{eq:12}
\end{equation}
for the part emitted  on the right at $\xi''$. The net current in the
ballistic interval  $[\xi',\xi'']$  is therefore
\begin{equation}
J(\xi', \xi'') = J_{\rm l}(\xi', \xi'') + J_{\rm r}(\xi', \xi'') = v_{\rm e}
N_{\rm c} e^{-\beta E_{\rm c}^{\rm m}(\xi',\xi'')} [e^{\beta E_{\rm F}(\xi')} -
e^{\beta E_{\rm F}(\xi'')}] \; ,
\label{eq:13}
\end{equation}
which is antisymmetric under the interchange of $\xi'$ and $\xi''$.

We now find for the total (net) electron current passing the point $\xi$ $(0 <
\xi < \sigma)$
\begin{eqnarray}
J(\xi) &=& p_{1}(\sigma) J(0,\sigma) + \int_{0}^{\xi }d\xi' p_{2}
(\sigma-\xi') J(\xi',\sigma)  \nonumber \\ &+& \int_{\xi}^{\sigma} d\xi''
p_{2}(\xi'') J(0,\xi'') + \int_{0}^{\xi}d\xi' \int_{\xi}^{\sigma} d\xi''
p_{3}(\xi''- \xi') J(\xi',\xi'') \; .
\label{eq:14}
\end{eqnarray}
Here, the first term on the right-hand side represents the net current of
electrons passing $\xi$ while moving ballistically between $0$ and $\sigma$
[which occurs with probability $p_{1}(\sigma)$], the second term refers to the
electrons in ballistic motion between any point $\xi'$ $(0 < \xi' < \xi)$ and
$\sigma$ [probability $p_{2}(\sigma - \xi') d\xi'$], etc.  In Fig.\
\ref{fig:2}, the four terms in expression (\ref{eq:14}) are represented by the
double-arrows labelled I to IV.

\begin{figure}
\epsfysize=12.0cm
\epsfbox[15 300 501 700]{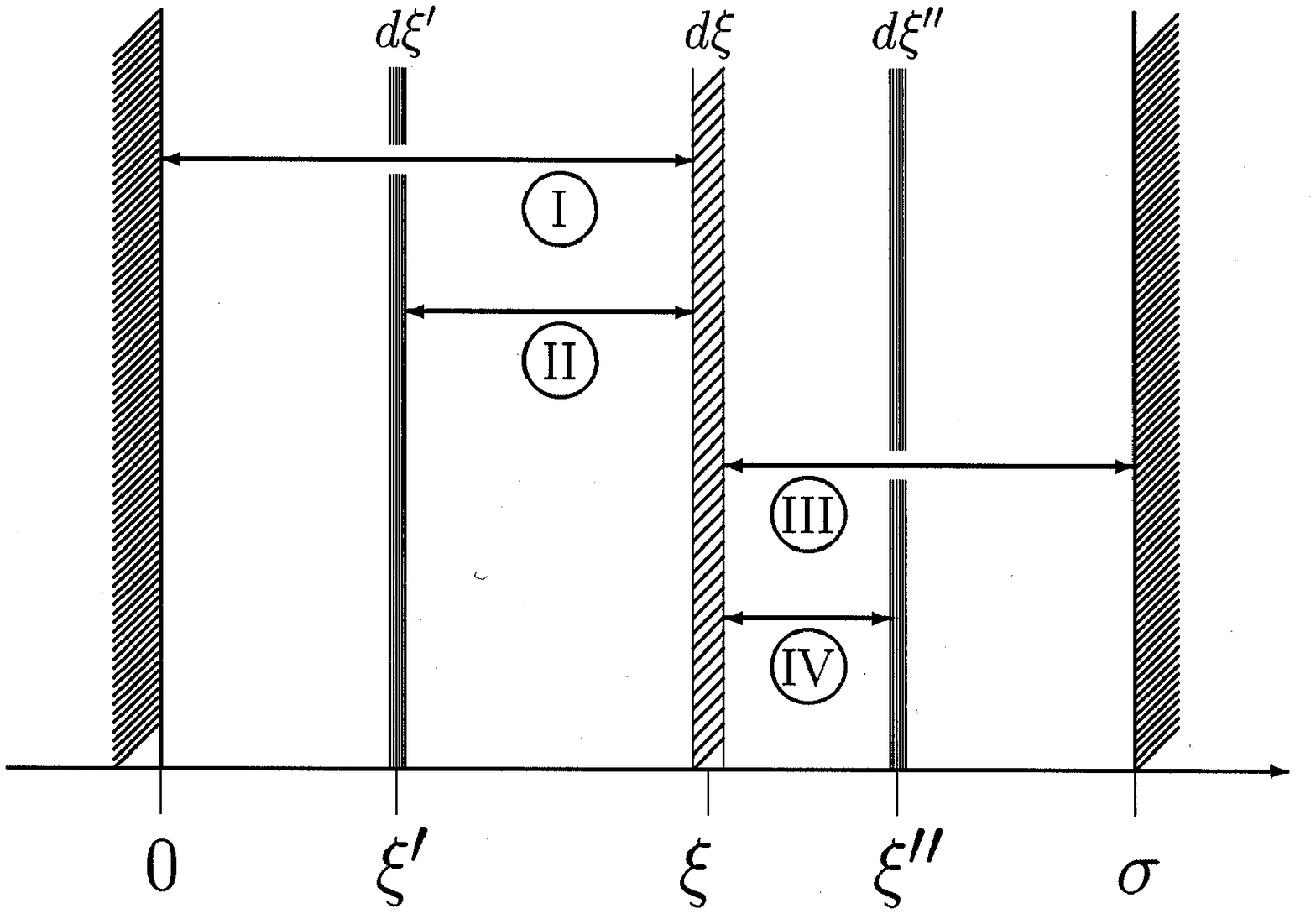}
\vspace*{4.0cm}
\caption{Diagram illustrating the four types of ballistic current entering into
the right-hand side of Eq.\ (\ref{eq:15}).}
\label{fig:3}
\end{figure}

We call the current $J(\xi)$ the {\em thermo-ballistic current}, since it is
the sum of weighted contributions of ballistic currents emitted and absorbed at
the contacts at $\xi = 0$ and $\xi = \sigma$, or at the local equilibration
(``thermalization'') points $\xi'$ and $\xi''$ inside the sample. For analyzing
the properties of this current, we consider its (positive or negative)
increment over the infinitesimal interval $[\xi, \xi+d\xi]$,
\begin{eqnarray}
dJ(\xi) = J(\xi+d\xi)-J(\xi) &=& [dJ(\xi)/d\xi] d\xi \nonumber \\ &
= &  - \left[ p_{2}(\xi) J(0,\xi) + \int_{0}^{\xi} d\xi' p_{3}(\xi-\xi')
J(\xi',\xi) \right] d\xi \nonumber \\[0.3cm] &\ & + \left[ \rule{0mm}{6mm}
p_{2}(\sigma - \xi) J(\xi,\sigma) + \int_{\xi}^{\sigma} d\xi''
p_{3}(\xi''-\xi) J(\xi,\xi'') \right] d\xi \; .
\label{eq:15}
\end{eqnarray}
The increment of the thermo-ballistic current is caused by the net influx
supplied to it via the emission and absorption (equilibration) processes
occurring in the interval $[\xi,\xi+d\xi]$, as depicted in Fig.\ \ref{fig:3}.
The net contribution of emission and absorption in that interval to the
thermo-ballistic current in the interval $[0,\xi]$ (indicated by the
double-arrow labelled I)is represented by the first term in the first pair of
brackets in expression (\ref{eq:15}), the net contribution to the
thermo-ballistic current in the interval $[\xi',\xi]$ (double-arrow II) by the
second term, and similarly for the two terms in the second pair of brackets
(double-arrows III and IV).

In order to compensate the local imbalance between emitted and absorbed
currents at the equilibration points, we introduce a ``background current''
$J^{\rm back}(\xi)$, in such a way that the total current $J^{\rm tot}(\xi)$,
defined as the sum of the thermo-ballistic current (\ref{eq:14}) and the
background current, is conserved,
\begin{equation}
J^{\rm tot}(\xi) = J(\xi) + J^{\rm back}(\xi) = J = {\rm const.}
\label{eq:16}
\end{equation}
This current is to be identified with the physical current.  Since, by its
nature, the background current is confined within the sample, it must vanish
when averaged over the latter,
\begin{equation}
\int_{0}^{\sigma} \frac{d\xi}{\sigma} \, J^{\rm back}(\xi) = 0 \; .
\label{eq:17}
\end{equation}
Now, averaging Eq.\ (\ref{eq:16}) over the sample and using Eq.\ (\ref{eq:17}),
we derive the condition of ``global conservation'' for the thermo-ballistic
current,
\begin{equation}
J = \int_{0}^{\sigma} \frac{d\xi}{\sigma} \, J(\xi) \; .
\label{eq:18}
\end{equation}
This relation expresses the physical current $J$ in terms of the
thermo-ballistic current $J(\xi)$; it will be the basis for setting up a
formalism which allows us to obtain the current-voltage characteristic, i.e., a
relation between the current $J$ and the voltage bias $V$.

\section{Current-voltage characteristic}

\subsection{General form of the current-voltage characteristic}

The thermo-ballistic current $J(\xi)$ is expressed by relation (\ref{eq:14}) in
terms of the two-variable function $J(\xi',\xi'')$.  In order to obtain an
expression containing a function that depends on the single variable $\xi$
only, we use Eq.\ (\ref{eq:13}) to write the current $J(\xi',\xi'')$ in the
ballistic interval $[\xi',\xi'']$ in terms of the ballistic current $J(0,\xi)$
in the interval $[0,\xi]$,
\begin{equation}
J(\xi',\xi'') = \gamma (\xi',\xi'') J(0,\xi'') - \gamma (\xi'',\xi') J(0,\xi')
\; ,
\label{eq:19}
\end{equation}
with
\begin{equation}
J(0, \xi)  = v_{\rm e} N_{\rm c} e^{-\beta E_{\rm c}^{\rm m}(0,\xi)}
[e^{\beta E_{\rm F}(0)} - e^{\beta E_{\rm F}(\xi)}] \; ,
\label{eq:19a}
\end{equation}
and the ``profile function'' $\gamma(\xi',\xi'')$ defined as
\begin{equation}
\gamma(\xi',\xi'') = e^{-\beta [E_{\rm c}^{\rm m}(\xi',\xi'') - E_{\rm
c}^{\rm m}(0,\xi'')]} \; .
\label{eq:20}
\end{equation}
The latter function is determined by the band edge profile $E_{\rm c}$, which
actually is a universal function of the coordinate $x$, independent of the mean
free path $l$, i.e., $E_{\rm c} = E_{\rm c}(x)$.  Therefore, as a function of
$x'$ and $x''$, $\gamma = \gamma(x',x'';[E_{\rm c}])$ is also independent of
$l$.  For notational convenience, {\em we continue to use the shorthand
notation} $E_{\rm c}(\xi)$ and $\gamma(\xi',\xi'')$ to represent the functions
$E_{\rm c}(x) = E_{\rm c}(l\xi)$ and $\gamma(x',x'';[E_{\rm c}]) =
\gamma(l\xi',l\xi'';[E_{\rm c}])$.  We note that the profile $E_{\rm c}(x)$ is
{\em indirectly} connected with $l$ via the dependence of both $E_{\rm c}$ and
$l$ on the doping concentration in the sample \cite{aro82}.

With the use of Eq.\ (\ref{eq:19}), expression (\ref{eq:14}) for the
thermo-ballistic current now becomes
\begin{eqnarray}
\frac{J(\xi)}{J} &=& \left[ p_{1}(\sigma)  + \int_{0}^{\xi }d\xi' p_{2}
(\sigma-\xi') \gamma(\xi',\sigma) \right] \chi(\sigma) \nonumber \\ &-&
\int_{0}^{\xi} d\xi' \left[ p_{2}(\sigma - \xi') \gamma(\sigma,\xi')  +
\int_{\xi}^{\sigma} d\xi'' p_{3}(\xi''- \xi')  \gamma(\xi'',\xi')
\right] \chi(\xi') \nonumber \\ &+& \int_{\xi}^{\sigma} d\xi' \left[
p_{2}(\xi')  + \int_{0}^{\xi} d\xi'' p_{3}(\xi'- \xi'') \gamma(\xi'',\xi')
\right] \chi(\xi') \; ,
\label{eq:21}
\end{eqnarray}
where we have introduced the (dimensionless) function
\begin{equation}
\chi(\xi) =   \frac{1}{J} \, J(0,\xi)  \; ,
\label{eq:25}
\end{equation}
which is defined for $0 \leq \xi \leq \sigma$.  If we assume, without loss of
generality, that $E_{\rm F}(0) > E_{\rm F}(\sigma)$, the physical current $J$
flows from left to right, $J > 0$, and the same holds for the ballistic current
between the contacts, $J(0,\sigma) > 0$ [cf.\ Eq.\ (\ref{eq:19a})].  Owing to
the effect of scattering, the physical current must be smaller than the
ballistic current, so that we have
\begin{equation}
\chi(\sigma) > 1 \; .
\label{eq:25b}
\end{equation}
The quantity $\widetilde{\chi}$ defined as
\begin{equation}
\widetilde{\chi} = \chi(\sigma)
\label{eq:25a}
\end{equation}
is the central quantity in the {\em current-voltage characteristic} following
from Eqs.\ (\ref{eq:19a}) [for $\xi = \sigma$] and (\ref{eq:25}),
\begin{equation}
j = -ev_{\rm e}N_{\rm c}\, e^{-\beta E_{\rm p}} \, \frac{1}{\widetilde{\chi}}
\, (1 - e^{- \beta eV}),
\label{eq:36}
\end{equation}
where $j = -eJ$ is the charge current, $E_{\rm p} = E_{\rm c}^{\rm m}(0,\sigma)
- E_{\rm F}(0)$ is the overall barrier height along the whole sample, and $V =
[E_{\rm F}(0) - E_{\rm F}(\sigma)]/e$ is the voltage bias between the contacts
at the ends of the sample.  For $\beta e V \ll 1$, Eq.\ (\ref{eq:36}) reduces
to $j = - V/R$, where the quantity $R$ (resistance times area of the sample) is
seen to be proportional to $\widetilde{\chi}$,
\begin{equation}
R = (e^{2} v_{\rm e} N_{\rm c} \beta)^{-1} \, e^{\beta E_{\rm p}} \,
\widetilde{\chi} \; .
\label{eq:38}
\end{equation}
We therefore call $\widetilde{\chi}$ the {\em reduced resistance}, and the
function $\chi(\xi)$ the {\em resistance function}.  We will see below that, in
general, $\widetilde{\chi}$ is not simply given by Eq.\ (\ref{eq:25a}) but must
be symmetrized with respect to the profile $E_{\rm c}$.

Performing the integration on Eq.\ (\ref{eq:21}) according to Eq.\
(\ref{eq:18}), we find
\begin{equation}
\sigma - f(\sigma) \chi(\sigma) + \int_{0}^{\sigma} d\xi \, {\cal
K}(\sigma,\xi)\, \chi(\xi) = 0 \; ,
\label{eq:26}
\end{equation}
with the coefficient function
\begin{equation}
f(\xi) = \xi p_{1}(\xi) + \int_{0}^{\xi} d\xi'
(\xi - \xi')  p_{2}(\xi - \xi') \gamma (\xi',\xi)
\label{eq:23}
\end{equation}
and the kernel
\begin{equation}
{\cal K}(\xi,\xi') = - \xi' p_{2}(\xi') + (\xi - \xi') p_{2}(\xi - \xi')
\gamma(\xi,\xi')  + \int_{0}^{\xi} d\xi''(\xi'' -\xi')p_{3}(|\xi'-\xi''|)
\gamma(\xi'',\xi')
\label{eq:24}
\end{equation}
$(\xi' \leq \xi)$.  Equation (\ref{eq:26}) is a condition on the function
$\chi(\xi)$, required by Eq.\ (\ref{eq:18}).  In the following, it will be
reinterpreted as an integral equation for $\chi(\xi)$, allowing us to determine
the reduced resistance $\widetilde{\chi} = \chi(\sigma)$.

\subsection{Determination of the resistance function}

The coefficient function $f(\xi)$ and the kernel ${\cal K}(\xi,\xi')$ depend on
the mean free path $l$ and the band edge profile $E_{\rm c}$ of the system via
the profile function $\gamma(\xi',\xi'') \equiv \gamma(l\xi',l\xi'';[E_{\rm
c}])$ [cf.\ Eq.\ (\ref{eq:20})],
\begin{equation}
f(\xi) = f(\xi;l,[E_{\rm c}]) \; ,
\label{eq:32}
\end{equation}
\begin{equation}
{\cal K}(\xi,\xi') = {\cal K}(\xi,\xi';l,[E_{\rm c}]) \; .
\label{eq:33}
\end{equation}
Apart from the implicit dependence on the profile contained in $f$ and ${\cal
K}$, Eq.\ (\ref{eq:26}) contains the position $\sigma = S/l$ of the right
contact explicitly.  We now allow this quantity to become a variable which may
assume any value between $\sigma = 0^{+}$ and $\sigma = S/l$.  Accordingly, by
replacing in Eq.\ (\ref{eq:26}) the quantity $\sigma$ with the variable $\xi$,
we obtain a Volterra-type integral equation for the resistance function
$\chi(\xi)$,
\begin{equation}
\xi - f(\xi) \chi(\xi) + \int_{0}^{\xi} d\xi' \, {\cal K}(\xi,\xi')\,
\chi(\xi') = 0 \; .
\label{eq:26a}
\end{equation}
By solving Eq.\ (\ref{eq:26a}) in the range $0 < \xi \leq \sigma$, we can
determine the function $\chi(\xi)$ which satisfies condition (\ref{eq:26}).

As a preliminary to the calculation of the function $\chi(\xi)$, we look at the
behavior of that function for $\xi \rightarrow 0$.  We use $\gamma(0,0) = 1$ in
Eqs.\ (\ref{eq:23}) and (\ref{eq:24}), and expand the right-hand sides of the
resulting expressions for $f(\xi)$ and ${\cal K}(\xi,\xi')$ [cf.\ Eqs.\
(\ref{eq:52}) and (\ref{eq:53}) below] to first order in $\xi$ and $\xi'$.
With the help of Eq.\ (\ref{eq:7}), we find $f(\xi) \approx \xi$ and ${\cal
K}(\xi,\xi') \approx p_{2}(0) (\xi - 2 \xi')$ for $\xi,\xi' \rightarrow 0$,
so that Eq.\ (\ref{eq:26a}) yields
\begin{equation}
\chi(0^{+}) = 1 \; .
\label{eq:28}
\end{equation}
Therefore, the function $\chi(\xi)$ is discontinuous at $\xi = 0$, since by
Eqs.\ (\ref{eq:19a}) and (\ref{eq:25}), $\chi(0) = 0$. Equation (\ref{eq:28})
defines the initial value for solving Eq.\ (\ref{eq:26a}). It implies
\begin{equation}
J(0,0^{+}) = J  \; ;
\label{eq:29}
\end{equation}
since $J \neq 0$, we have $E_{\rm F}(0^{+}) \neq E_{\rm F}(0)$ [cf.\ Eq.\
(\ref{eq:19a})], i.e., the quasi-Fermi level for $\xi = 0^{+}$ differs from the
Fermi level in the contact at $\xi = 0$ by an amount determined by Eq.\
(\ref{eq:28}).

Equation (\ref{eq:26a}) is the key result of the formal development of this
paper.  It is a linear equation as long as the band edge profile $E_{\rm
c}(\xi)$, and consequently the profile function $\gamma(\xi',\xi'')$, can be
considered given, as in the case of zero bias.  Otherwise, the profile function
will depend on the solution $\chi(\xi)$ which, via the quasi-Fermi level
$E_{\rm F}(\xi)$, enters the Poisson equation for $E_{\rm c}(\xi)$.  Equation
(\ref{eq:26a}) then becomes a nonlinear equation which must be solved
self-consistently together with the Poisson equation.

Once the function $\chi(\xi)$ has been determined for $0 \leq \xi \leq \sigma$,
we can obtain the quasi-Fermi level $E_{\rm F}(\xi)$ as a function of the
parameter $J$ by using  Eqs.\ (\ref{eq:19a}) and (\ref{eq:25}):
\begin{equation}
E_{\rm F}(\xi) = E_{\rm F}(0) + \frac{1}{\beta}\, \ln \left[ 1 -
\frac{J}{v_{\rm e} N_{\rm c}} \chi(\xi) \, e^{-\beta [E_{\rm F}(0) -
E_{\rm c}^{\rm m}(0,\xi)]} \right]
\; .
\label{eq:39a}
\end{equation}
Eliminating in this expression the current $J = -j/e$ with the help of the
current-voltage characteristic (\ref{eq:36}), we can also write  $E_{\rm
F}(\xi)$ as a function of the parameter $V$:
\begin{equation}
E_{\rm F}(\xi) = E_{\rm F}(0) + \frac{1}{\beta}\, \ln \left[ 1 -
\frac{\chi(\xi)}{\chi(\sigma)} \, e^{-\beta [E_{\rm c}^{\rm
m}(0,\sigma) - E_{\rm c}^{\rm m}(0,\xi)]} \, (1 - e^{- \beta eV}) \right]
\; .
\label{eq:39}
\end{equation}
The thermo-ballistic current $J(\xi)$ inside the sample is found from Eq.\
(\ref{eq:21}), with the physical current $J$ entering as an overall factor
only.

\subsection{Interpretation}

Since the resistance function approaches unity as $\xi \rightarrow 0^{+}$, it
is appropriate to write it (for $\xi > 0$) in the form
\begin{equation}
\chi(\xi) = 1 + \kappa(\xi) \xi \; .
\label{eq:30}
\end{equation}
From the examples presented in Sec.\ V, it emerges that $\kappa(\xi) > 0$,
that $\kappa(0)$ is finite, and that in the limit $\xi \rightarrow \infty$, the
function $\kappa(\xi)$ tends toward a finite value.

The first term on the right-hand side of Eq.\ (\ref{eq:30}), i.e., the limiting
value of $\chi(\xi)$ for $\xi = x/l \rightarrow 0^{+}$, represents the purely
ballistic part of the resistance if this limit is thought to have been obtained
for $l \rightarrow \infty$ at arbitrary fixed $x$.  On the other hand, if we
let $x \rightarrow 0^{+}$ at arbitrary fixed $l$, this term is to be
interpreted as a contact resistance:\ from Eqs.\ (\ref{eq:19a}) and
(\ref{eq:29}), we find that a finite (but small compared with $k_{\rm B}T/e$)
voltage difference $V(0^{+}) = [E_{\rm F}(0) - E_{\rm F}(0^{+})]/e$ across the
interval $[0,0^{+}]$ is associated with a {\em finite} conductance per unit
area
\begin{equation}
G_{\rm C} = e \left| \frac{J}{V(0^{+})} \right| = e \left|
\frac{J(0,0^{+})}{V(0^{+})} \right| = e^{2} v_{\rm e} N_{\rm c} \beta e^{-\beta
[E_{\rm c}(0) - E_{\rm F}(0)]}
\label{eq:31}
\end{equation}
(recall $j = -eJ$), although apparently there is no hindrance to the current in
that interval.  This conductance is the inverse of the contact resistance
located in an ideal, ``reflectionless'' contact \cite{lan87,dat95,imr99} at
$\xi = 0$; it is the nondegenerate semiconductor analogue of the contact
conductance  for a three-dimensional degenerate electron conductor
at zero temperature \cite{bee91}, $G_{\rm C} = (2e^{2}/h)(k_{\rm F}^{2}/4\pi)$,
where $k_{\rm F}^{2} = 2m^{*}[E_{\rm F}(0) - E_{\rm c}(0)]/\hbar^{2}$.  The
contact resistance is located inside the left contact because the resistance
function $\chi(\xi)$ is associated, via Eqs.\ (\ref{eq:19a}) and (\ref{eq:25}),
with a quasi-Fermi level $E_{\rm F}(\xi)$ determining the current injected from
the right, which enters the left contact without reflection \cite{dat95}.

Turning now to the interpretation of the thermo-ballistic current $J(\xi)$, we
restrict ourselves, for the sake of transparency, to the special case of
{\em constant} band edge profile, $E_{\rm c}(x) = E_{\rm c}(0)$.  Then the
current-voltage characteristic reads
\begin{equation}
j = -ev_{\rm e}N_{\rm c}\, e^{-\beta [E_{\rm c}(0) - E_{\rm F}(0)]} \,
\frac{1}{\chi(\sigma)} \, (1 - e^{- \beta eV})
\label{eq:54}
\end{equation}
[cf.\ Eq.\ (\ref{eq:36})], where $\chi(\sigma)$ is a universal function of
$\sigma$.  The quasi-Fermi level is given by
\begin{equation}
E_{\rm F}(\xi) = E_{\rm F}(0) + \frac{1}{\beta}\, \ln \left[ 1 -
\frac{\chi(\xi)}{\chi(\sigma)} \, (1 - e^{- \beta eV}) \right]
\label{eq:55}
\end{equation}
[cf.\ Eq.\ (\ref{eq:39})], and the thermo-ballistic current $J(\xi)$ is
obtained from Eq.\ (\ref{eq:21}) as
\begin{equation}
\frac{J(\xi)}{J} =  p_{1}(\sigma-\xi) \, \chi(\sigma) -
\int_{0}^{\sigma} d\xi' \, p_{2}(|\xi-\xi'|) \, {\rm sgn}(\xi-\xi') \,
\chi(\xi') \; ,
\label{eq:56}
\end{equation}
where use has been made of Eq.\ (\ref{eq:7}).

Equation (\ref{eq:56}) may be rewritten in the form
\begin{equation}
\frac{J(\xi)}{J} =  p_{1}(\xi) \, \chi(0^{+}) + \int_{0}^{\sigma} d\xi' \,
p_{1}(|\xi-\xi'|) \, \frac{d}{d\xi'} \chi(\xi') \; .
\label{eq:56a}
\end{equation}
Introducing the ``internal voltage difference'' across the interval $[0,\xi]$,
\begin{equation}
V(\xi) = [E_{\rm F}(0) - E_{\rm F}(\xi)]/e \; ,
\label{eq:56b}
\end{equation}
and the thermo-ballistic charge current $j(\xi) = -e J(\xi)$, we now have,
using Eqs.\ (\ref{eq:28}), (\ref{eq:54}), and (\ref{eq:55}),
\begin{equation}
j(\xi) = \frac{p_{1}(\xi)}{\chi(\sigma)} \, j^{\rm ball} - \int_{0}^{\sigma}
d\xi' \, {\cal C}(\xi,\xi') \, \frac{d}{d\xi'} V(\xi') \; .
\label{eq:56d}
\end{equation}
Here,
\begin{equation}
j^{\rm ball}  = - e  v_{\rm e} N_{\rm c} \, e^{-\beta  [E_{\rm c}(0) - E_{\rm
F}(0)]} \; (1 - e^{-\beta e V})
\label{eq:56f}
\end{equation}
is the current in the ballistic limit, and
\begin{equation}
{\cal C}(\xi,\xi') = \frac{1}{2 \langle l \rangle} \; p_{1}(|\xi-\xi'|) \;
c(\xi') \; ,
\label{eq:56e}
\end{equation}
where $c(\xi) = e \mu n(\xi)$ is the (local) conductivity, with the mobility
$\mu = 2e v_{\rm e} \beta \langle l \rangle$.  In expression (\ref{eq:56d}) for
the thermo-ballistic current at position $\xi$, the first term represents the
ballistic component which enters with weight factor $p_{1}(\xi)/\chi(\sigma)$.
The second term represents the diffusive component, driven by the internal
voltage gradient $- dV(\xi)/d\xi$, with a {\em nonlocal} conductivity ${\cal
C}(\xi,\xi')$.  In the ballistic limit, we have $j(\xi) = j = j^{\rm ball}$,
while in the diffusive limit ${\cal C}(\xi,\xi') = {\cal C}(x/l,x'/l) \propto
\delta(x - x')$, so that
\begin{equation}
j(x) = j = - c(x) \; \frac{d}{dx} V(x) \; .
\label{eq:56g}
\end{equation}
In the Appendix, these limits will be worked out for the case of arbitrary band
edge profile.

The starting point of the development in Sec.\ IV.A has been to rewrite the
current $J(\xi',\xi'')$ in terms of the current $J(0,\xi)$ containing the
quasi-Fermi level $E_{\rm F}(\xi)$ at the right end of the interval $[0,\xi]$
extending between the left contact at $\xi = 0$ and the position $\xi$.
However, in the formalism of Sec.\ III no preference is given to either
contact. Therefore, we could just as well have chosen to write the current
$J(\xi',\xi'')$ in terms of the current $J(\xi,\sigma)$ containing the
quasi-Fermi level at the left end of the interval $[\xi, \sigma]$ extending
between the position $\xi$ and the right contact at $\xi = \sigma$.  The
modifications caused by this alternative choice are the following.

The function $\chi(\xi;[E_{\rm c}])$ is replaced with the function
$\chi_{1}(\xi;[E_{\rm c}]) \equiv J(\xi,\sigma;[E_{\rm c}])/J = \chi(\sigma -
\xi;[E_{\rm c}^{*}])$, and $J(\xi;[E_{\rm c}])$ is replaced with $J(\sigma -
\xi;[E_{\rm c}^{*}])$; here, $E_{\rm c}^{*}$ symbolizes the reversed profile,
$E_{\rm c}^{*}(\xi) = E_{\rm c}(\sigma - \xi) $. The function
$\chi_{1}(\xi;[E_{\rm c}])$ is discontinuous at the right contact $\xi =
\sigma$, and the contact resistance is located there.  Since the quasi-Fermi
level $E_{\rm F}(\xi)$ and the thermo-ballistic current $J(\xi)$ depend on the
chosen location of the contact resistance, they are to be considered as
auxiliary functions that have no immediate physical meaning.  They are,
nevertheless, of great help in the interpretation of the results, and will be
considered in this light in the discussion of the examples of Sec.~V.  On the
other hand, the reduced resistance $\widetilde{\chi}$, which is a physical
quantity entering the current-voltage characteristic, must be unique.  This
requirement can be met in a natural way by symmetrization:\ since we have
$\chi_{1}(0;[E_{\rm c}]) = \chi(\sigma;[E_{\rm c}^{*}])$, we take
$\widetilde{\chi}$ in Eq.\ (\ref{eq:36}) in the symmetrized form
\begin{equation}
\widetilde{\chi} =\frac{1}{2} \left\{ \chi(\sigma;[E_{\rm c}]) +
\chi_{1}(\sigma;[E_{\rm c}]) \right\} = \frac{1}{2} \left\{ \chi(\sigma;[E_{\rm
c}]) + \chi(\sigma;[E_{\rm c}^{*}]) \right\} \; ,
\label{eq:36a}
\end{equation}
which replaces Eq.\ (\ref{eq:25a}). For constant profile, in which case $E_{\rm
c}^{*} = E_{\rm c}$, Eq.\ (\ref{eq:36a}) reduces to Eq.\ (\ref{eq:25a}). In
analogy to Eq.\ (\ref{eq:30}), it is convenient to write
\begin{equation}
\widetilde{\chi} = 1 + \widetilde{\kappa} \sigma \; ,
\label{eq:36bb}
\end{equation}
where
\begin{equation}
\widetilde{\kappa} =  \frac{1}{2} \left\{ \kappa(\sigma;[E_{\rm c}]) +
\kappa(\sigma;[E_{\rm c}^{*}]) \right\} .
\label{eq:36c}
\end{equation}
Since $\sigma = S/l$ and the sample length $S$ is included in the definition of
the profile $E_{\rm c}$, the quantity $\widetilde{\chi}$ (and thus also
$\widetilde{\kappa}$) is solely a function of $l$ and a (symmetrized)
functional of $E_{\rm c}$.

\subsection{The degenerate case for zero bias}

In the degenerate case, a simple treatment is possible only in the limit of
zero bias.  In Eq.\ (\ref{eq:14}), we must replace the current $J(\xi', \xi'')$
of Eq.\ (\ref{eq:13}) with
\begin{equation}
J^{\rm d}(\xi', \xi'') = v_{\rm e} N_{\rm c} \beta \, \ln(1 + e^{-\beta [E_{\rm
c}^{\rm m}(\xi',\xi'') - E_{\rm F}(0)]}) \, [E_{\rm F}(\xi') - E_{\rm
F}(\xi'')] \; ,
\label{eq:46}
\end{equation}
where $E_{\rm F}(\xi')$ and $E_{\rm F}(\xi'')$ deviate infinitesimally from
$E_{\rm F}(0)$.  Introducing, in analogy to Eq.~(\ref{eq:25}), the resistance
function
\begin{equation}
\chi^{\rm d}(\xi) =   \frac{1}{J} \, J^{\rm d}(0,\xi) \; ,
\label{eq:47}
\end{equation}
we write $J^{\rm d}(\xi', \xi'')$ in the form
\begin{equation}
J^{\rm d}(\xi',\xi'') = J[\gamma^{\rm d} (\xi',\xi'') \chi^{\rm d}(\xi'') -
\gamma^{\rm d}(\xi'',\xi') \chi^{\rm d}(\xi')] \; ,
\label{eq:48}
\end{equation}
where the profile function $\gamma^{\rm d} (\xi',\xi'')$ is defined as
\begin{equation}
\gamma^{\rm d}(\xi',\xi'') = \frac{\ln(1 + e^{-\beta [E_{\rm c}^{\rm
m}(\xi',\xi'') - E_{\rm F}(0)]})}{\ln(1 + e^{-\beta [E_{\rm c}^{\rm
m}(0,\xi'') - E_{\rm F}(0)]})}  \; .
\label{eq:49}
\end{equation}
We then find for $\chi^{\rm d}(\xi)$ an integral equation which is identical to
that for $\chi(\xi)$ [cf.\ Eqs.\ (\ref{eq:23}), (\ref{eq:24}), and
(\ref{eq:26a})], except that the function $\gamma(\xi',\xi'')$ is to be
replaced with $\gamma^{\rm d} (\xi',\xi'')$.  The zero-bias conductance per
unit area, $G = |j/V|_{V \rightarrow 0}$, is obtained as
\begin{equation}
G = e^{2} v_{\rm e} N_{\rm c} \beta \, \ln (1+ e^{-\beta E_{\rm p}}) \,
\frac{1}{\widetilde{\chi}^{\rm d}} \; ,
\label{eq:50}
\end{equation}
where
\begin{equation}
\widetilde{\chi}^{\rm d} = \frac{1}{2} \left\{ \chi^{\rm d}(\sigma;[E_{\rm
c}]) + \chi^{\rm d}(\sigma;[E_{\rm c}^{*}]) \right\}
\label{eq:36b}
\end{equation}
by analogy with Eq.\ (\ref{eq:36a}).

\section{Illustrative examples}

In order to illustrate the formalism developed in the previous sections, we now
consider a number of simple examples.  We present results of calculations for
the central quantity appearing in the current-voltage characteristic, the
reduced resistance, as well as for the quasi-Fermi level and the
thermo-ballistic current.  In particular, we study the dependence of these
quantities on the ratio of mean free path to sample length and on the shape of
the band edge profile.  To assess the effect of the dimensionality of the
electron motion, we compare results for one-dimensional and three-dimensional
transport; considering the properties of the probability functions $p_{n}(\xi)$
[cf.\ Sec.\ II], we may expect the results for two-dimensional transport to
fall in between those for one- and three dimensional motion throughout.

We first examine the case of constant band edge profile. As an example for
non-constant profiles, we then consider profiles containing a single maximum;
this case includes profiles that rise or fall monotonically over the length of
the sample.  The profile function $\gamma(\xi',\xi'')$, which is given [cf.\
Eq.\ (\ref{eq:20})] in terms of the maximum $E_{\rm c}^{\rm m}(\xi',\xi'')$ of
the profile in the interval $[\xi',\xi'']$, can be reduced in all these cases
to an explicit function of the profile $E_{\rm c}(\xi)$ itself.  This in turn
allows the functions $f(\xi)$ and ${\cal K}(\xi,\xi')$ to be written in a more
explicit, simplified form.  For profiles more complicated than the ones
considered here, the profile function in general must be determined
numerically.

\subsection{Constant band edge profile}

In the case of constant conduction band edge profile (``flat band edge''), we
have $\gamma(\xi',\xi'') = 1$.  For the coefficient function $f$ and the kernel
${\cal K}$, we then find from Eqs.\ (\ref{eq:23}) and (\ref{eq:24}), using
Eqs.\ (\ref{eq:7}) and (\ref{eq:9}),
\begin{equation}
f(\xi) =  p_{0}(0) - p_{0}(\xi)
\label{eq:52}
\end{equation}
and
\begin{equation}
{\cal K}(\xi,\xi') = p_{1}(\xi')-p_{1}(\xi - \xi') \; ,
\label{eq:53}
\end{equation}
respectively.  As the functions $f$ and ${\cal K}$ are universal functions of
the scaled variables $\xi$ and $\xi'$, the solution $\chi(\xi)$ of Eq.\
(\ref{eq:26a}) will also be solely a function of $\xi$.

For {\em one-dimensional} motion, we find from Eqs.\ (\ref{eq:52}) and
(\ref{eq:53}), using Eq.\ (\ref{eq:57}),
\begin{equation}
f(\xi) =  1- e^{-\xi}
\label{eq:58}
\end{equation}
and
\begin{equation}
{\cal K}(\xi,\xi') = e^{-\xi'}-e^{-(\xi - \xi')} \; ,
\label{eq:59}
\end{equation}
respectively.  The solution of Eq.\ (\ref{eq:26a}) is then seen to be
\begin{equation}
\chi(\xi) = 1 + \xi/2 \; ,
\label{eq:60}
\end{equation}
which is valid in the range $0 < \xi \le \sigma$ [cf.\ Eq.\ (\ref{eq:28})],
and the function $\kappa(\xi)$ introduced in Eq.\ (\ref{eq:30}) assumes the
constant value $1/2$.

Specializing now Eqs.\ (\ref{eq:54})-(\ref{eq:56}) to the one-dimensional case,
we obtain for the current-voltage characteristic
\begin{equation}
j = -e v_{\rm e} N_{\rm c} \, e^{-\beta [E_{\rm c}(0) - E_{\rm F}(0)]} \,
\frac{1}{1 + \sigma/2} \; (1 - e^{- \beta eV})
\label{eq:61}
\end{equation}
(with $e v_{\rm e} N_{\rm c} = 2e/\beta h$), for the quasi-Fermi level
\begin{equation}
E_{\rm F}(\xi) = E_{\rm F}(0) + \frac{1}{\beta}\, \ln \left[ 1 -
\frac{1 + \xi/2}{1 + \sigma/2} \, (1 - e^{- \beta eV}) \right] \; ,
\label{eq:62}
\end{equation}
and for the thermo-ballistic current
\begin{equation}
\frac{J(\xi)}{J} =  1 + \smfrac{1}{2}(e^{-\xi} - e^{-(\sigma-\xi)}) \; .
\label{eq:63}
\end{equation}

For {\em three-dimensional} motion, we have from Eqs.\ (\ref{eq:52}) and
(\ref{eq:53}), using Eqs.\ (\ref{eq:70}) and (\ref{eq:71}),
\begin{equation}
f(\xi) = \frac{2}{3} - 2 E_{4}(\xi)
\label{eq:72}
\end{equation}
and
\begin{equation}
{\cal K}(\xi,\xi') = 2 [E_{3}(\xi') - E_{3}(\xi-\xi')] \; ,
\label{eq:73}
\end{equation}
respectively.  Then, solving Eq.\ (\ref{eq:26a}) numerically for the resistance
function $\chi(\xi)$, we obtain the reduced resistance $\widetilde{\chi} =
\chi(\sigma)$, which determines the current-voltage characteristic via
Eq.~(\ref{eq:54}).  The quasi-Fermi level and the thermo-ballistic current can
be calculated from Eqs.~(\ref{eq:55}) and (\ref{eq:56}), respectively.

\begin{figure}
\epsfysize=14.0cm
\epsfbox[66 206 503 623]{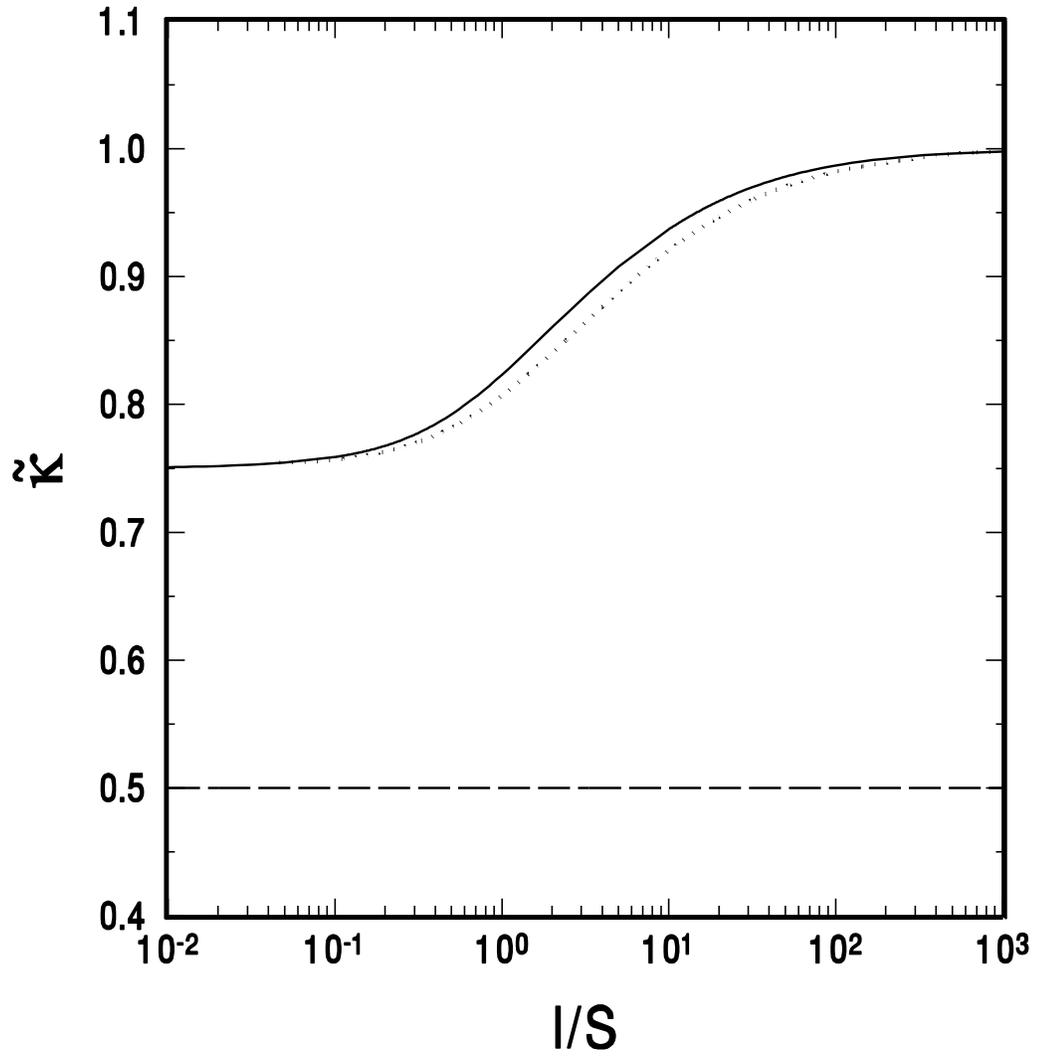}
\vspace*{2.0cm}
\caption{The quantity $\widetilde{\kappa}$ characterizing the reduced
resistance $\widetilde{\chi}$ for flat band edge profile, plotted as a function
of $l/S$.  Solid curve:\ three-dim\-ensional transport; long-dashed line:\
one-dimensional transport; dotted curve:\ result derived from the results of
Ref.\ \protect\cite{dej94}.}
\label{fig:4}
\end{figure}

In order to characterize the behavior of the reduced resistance
$\widetilde{\chi}$ for flat band edge profile, we show in Fig.\ \ref{fig:4} the
quantity $\widetilde{\kappa}$ [cf.\ Eq.\ (\ref{eq:36c})], plotted against
$l/S$.  For three-dim\-ensional transport, the curve for $\widetilde{\kappa}$
varies smoothly between the limits $l/S \ll 1$ (diffusive limit), where $\kappa
\rightarrow 3/4$, and $l/S \gg 1$ (ballistic limit), where $\kappa \rightarrow
1$; accordingly, there is a smooth transition of the current-voltage
characteristic from diffusive to ballistic behavior.  For one-dimensional
transport, we simply have $\widetilde{\kappa} = 1/2$ [cf.\ Eq.\ (\ref{eq:60})].
The reduced resistance is higher for three-dimensional than for one-dimensional
transport, owing to the smaller effective mean free path in the former case.
In fact, in the diffusive limit, we have $\widetilde{\chi} \rightarrow
\widetilde{\kappa} \, S/l$, and the ratio of the reduced resistances for
three-dimensional and one-dimensional transport is equal to 3/2, which is the
inverse of the ratio of the corresponding effective mean free paths (cf.\ Sec.\
II).  This result holds for any profile [cf.\ Eq.\ (\ref{eq:44})].  On the
other hand, in the ballistic limit the ratio of the quantities
$\widetilde{\kappa}$ for three-dimensional and one-dimensional transport is
equal to 2, not 3/2.  However, this has only little effect in the reduced
resistance because the contribution of $\widetilde{\kappa}$ in Eq.\
(\ref{eq:36bb}) is suppressed as the ballistic limit is approached.

Also shown in Fig.\ \ref{fig:4} is a curve for $\widetilde{\kappa}$ that has
been derived from the results for flat profile and three-dimensional transport
obtained in Ref.\ \cite{dej94} and that, while coinciding with our results in
the diffusive and ballistic limits, exhibits small deviations from our curve in
the range of intermediate $l/S$.  We return to a discussion of this feature in
Sec.\ VI.

\begin{figure}
\epsfysize=14.0cm
\epsfbox[66 206 503 623]{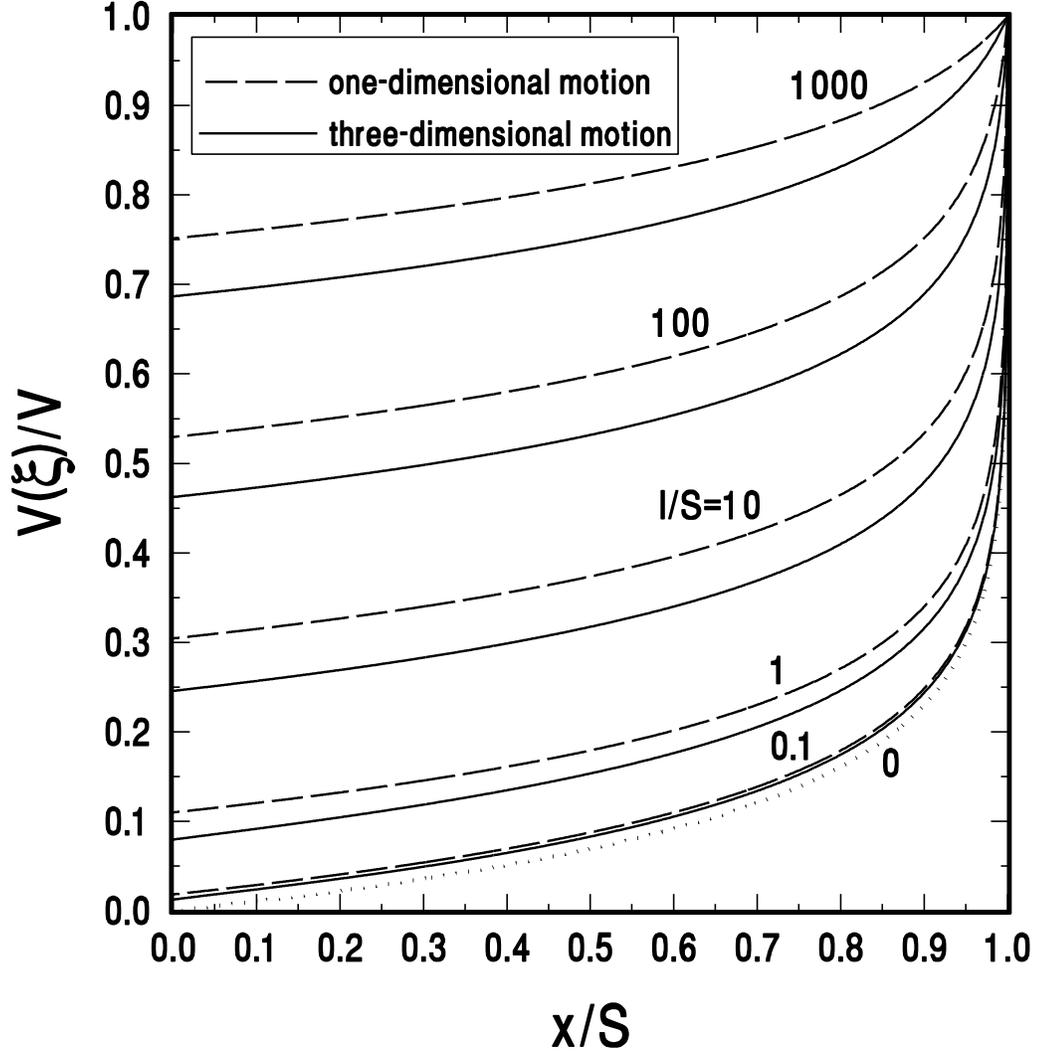}
\vspace*{2.0cm}
\caption{The normalized internal voltage difference $V(\xi)/V$ for one- and
three-dimensional transport across a flat band edge profile, plotted as a
function of the reduced coordinate $x/S = l\xi/S$ for various values of $l/S$
($\beta eV = eV/k_{\rm B}T = 10$).  The dotted curve labelled ``0'' corresponds
to $l/S \ll 1$ and has been calculated from Eq.\ (\ref{eq:73a}).}
\label{fig:5}
\end{figure}

To demonstrate the behavior of the quasi-Fermi level $E_{\rm F}(\xi)$, we show
in Fig.\ \ref{fig:5} the normalized internal voltage difference $V(\xi)/V$
[cf.\ Eq.\ (\ref{eq:56b})], plotted versus the reduced coordinate $x/S =
l\xi/S$. Calculated with $\beta eV = eV/k_{\rm B}T = 10$, the results of Fig.\
\ref{fig:5} are typical of the strong-bias regime.  The curve corresponding to
the diffusive limit $l/S \ll 1$ reflects the expression
\begin{equation}
E_{\rm F}(\xi) = E_{\rm F}(0) + \frac{1}{\beta} \, \ln \left[ 1 -
\frac{l\xi}{S} \, (1 - e^{-\beta e V}) \right]
\label{eq:73a}
\end{equation}
that follows by combining Eq.\ (\ref{eq:44}) [with $\widetilde{S} = S$] with
the integrated form of Eq.\ (\ref{eq:45}) and that is independent of the
dimensionality of the electron motion.  For $l/S > 0$, the curves have a
discontinuity at $x = 0$ [recall that $E_{\rm F}(0^{+}) \neq E_{\rm F}(0)$,
and hence $V(0^{+}) \neq V(0)$], which increases roughly proportional to
$\ln(l/S)$ up to $l/S \approx 1000$. This voltage gap is associated with the
contact resistance at $\xi = 0$. The latter amounts to the total resistance in
the ballistic limit, but its contribution to the voltage difference $V(\xi)$
becomes smaller and smaller as the diffusive limit is approached.  The results
for three-dimensional transport are below those for one-dimensional transport,
for the same reasons as above.  For values of $\beta e V$ smaller than the
chosen value of 10, all curves in Fig.\ \ref{fig:5} would be shifted upwards.

\begin{figure}
\epsfysize=14.0cm
\epsfbox[66 206 503 623]{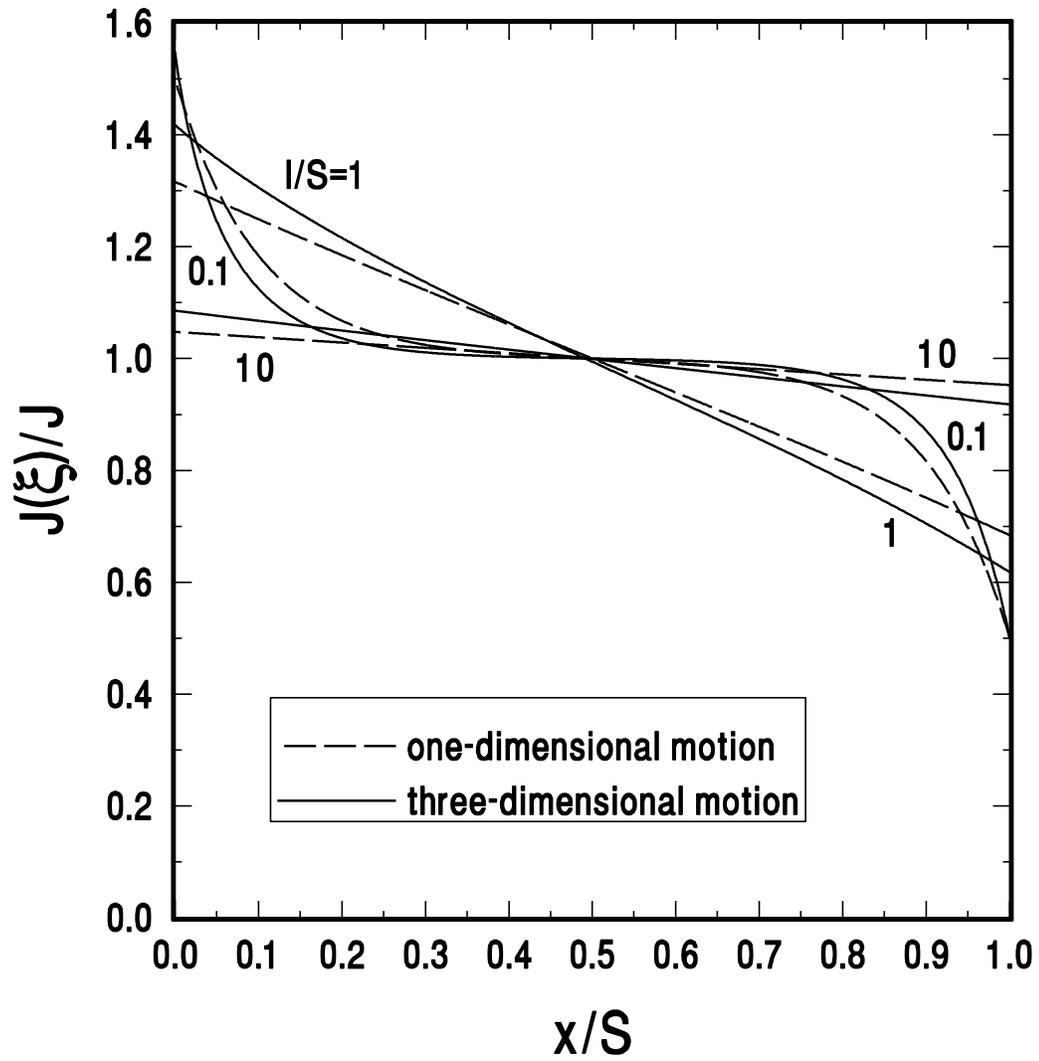}
\vspace*{2.0cm}
\caption{The normalized thermo-ballistic current $J(\xi)/J$ for one- and
three-dimensional transport across a flat band edge profile, plotted as a
function of the reduced coordinate $x/S = l\xi/S$ for various values of $l/S$.}
\label{fig:6}
\end{figure}

The thermo-ballistic current $J(\xi)$ for flat band edge profile, normalized to
the physical current $J$, is shown in Fig.\ \ref{fig:6} as a function of the
reduced coordinate $x/S = l\xi/S$.  The qualitative behavior of the curves for
small $x/S$ can be understood by reverting to Eq.\ (\ref{eq:56a}):\ with
$\chi(0^{+}) = 1$ [cf.\ Eq.\ (\ref{eq:28})] and $\chi'(\xi) > 0$ (as can be
inferred from the universal curve for $c_{0} = 0$ in Fig.\ \ref{fig:4}), we
have $J(\xi)/J > 1$ in the vicinity of $x = 0$.  For $x$ in the vicinity of
$S$, we cannot argue in this way; however, the drop of $J(\xi)/J$ below unity
in this range is clearly in agreement with Eq.\ (\ref{eq:18}).

For one-dimensional transport [cf.\ Eq.\ (\ref{eq:63})], the function
$[J(\xi)/J] - 1$ is strictly antisymmetric with respect to an inversion of the
reduced coordinate at $x/S = 1/2$.  This symmetry is (weakly) violated in the
three-dimensional case.  In conformance with the discussion of Sec.\ IV.C, the
thermo-ballistic current in the ballistic limit, $l/S \gg 1$, tends to become
equal to the physical current over the whole sample, $J(\xi)/J \rightarrow 1$,
independent of the dimensionality of the electron motion.  The same holds in
the diffusive limit $l/S \ll 1$, except for the vicinity of the left and right
ends of the sample, where $J(x \rightarrow 0)/J = 3/2$ and $J(x/S \rightarrow
1)/J = 1/2$, respectively, for one-dimensional transport, and $J(x \rightarrow
0)/J = 1.5485$ and $J(x/S \rightarrow1)/J = 1/2$ for three-dimensional
transport.

This behavior of the thermo-ballistic current is in accordance with the
comments made in connection with the internal voltage difference.  In the
ballistic limit, the total resistance is equal to the contact resistance, and
thermo-ballistic, ballistic, and physical current all coincide.  In the
diffusive limit, the effect of the contact resistance is suppressed, and the
thermo-ballistic current becomes entirely diffusive and equal to the physical
current.  However, the diffusive component is defined only inside the sample,
and the discontinuities of $J(\xi)$ at the contacts for $l/S \rightarrow 0$
reflect the residual influence of the contact resistance.

\subsection{Non-constant band edge profile}

For band edge profiles $E_{\rm c}(\xi)$ that are not constant over the length
of the sample, the reduced resistance $\widetilde{\chi}(l,[E_{\rm c}])$ is to be
calculated from the general form of Eq.\ (\ref{eq:26a}), with the coefficient
function $f(\xi) = f(\xi;[E_{\rm c}])$ and the kernel ${\cal K}(\xi,\xi') =
{\cal K}(\xi,\xi';[E_{\rm c}])$ depending on $E_{\rm c}(\xi)$ via the profile
function $\gamma(\xi',\xi'')$.

Examples with non-constant profiles allow us to study the interplay of the
effects of the profile structure and of the magnitude of the mean free path.
We investigate the behavior of the internal voltage difference, of the
thermo-ballistic current, and of the reduced resistance.  In the
current-voltage characteristic (\ref{eq:36}), of course, the main effect of the
profile generally is the barrier effect represented by the factor $e^{- \beta
E_{\rm p}}$.

Here, we consider the case of profiles exhibiting a single maximum at $S_{1}$
in the interval $[0,S]$, i.e., at $\xi = \sigma_{1} = S_{1}/l$.  From Eq.\
(\ref{eq:20}), we find for the corresponding profile function
\begin{eqnarray}
\gamma^{(1)}(\xi',\xi'') &=& [ \widetilde{\gamma}(\max(\xi',\xi''))
\Theta(\sigma_{1}-\xi') + \widetilde{\gamma}(\sigma_{1})
\Theta(\xi'-\sigma_{1})]/[\widetilde{\gamma}(\xi'')]
\Theta(\sigma_{1}-\xi'') \nonumber \\ &+&
[\widetilde{\gamma}(\sigma_{1}) \Theta(\sigma_{1}-\xi')
+ \widetilde{\gamma}(\min(\xi',\xi''))
\Theta(\xi' - \sigma_{1})]/[\widetilde{\gamma}(\sigma_{1})]
\Theta(\xi''-\sigma_{1}) \; , \label{eq:82}
\end{eqnarray}
where we have defined
\begin{equation}
\widetilde{\gamma}(\xi) = e^{-\beta E_{\rm c}(\xi)} \; .
\label{eq:75}
\end{equation}
For the coefficient function $f^{(1)}(\xi)$ related to
$\gamma^{(1)}(\xi',\xi'')$, we obtain from Eq.\ (\ref{eq:23})
\begin{equation}
f^{(1)}(\xi) = f^{(<)}(\xi) \Theta(\sigma_{1} - \xi) + f^{(>)}(\xi)
\Theta(\xi-\sigma_{1})  \; ,
\label{eq:83}
\end{equation}
where
\begin{equation}
f^{(<)}(\xi) =  p_{0}(0) - p_{0}(\xi)
\label{eq:76}
\end{equation}
[note that $f^{(<)}(\xi)$ does not depend on the band edge profile and agrees
with the coefficient function $f(\xi)$ for flat profile given by Eq.\
(\ref{eq:52})] and
\begin{eqnarray}
f^{(>)}(\xi) &=& p_{0}(\xi - \sigma_{1}) - p_{0}(\xi) + (\xi - \sigma_{1})
p_{1}(\xi - \sigma_{1}) \nonumber \\ &+& [1/\widetilde{\gamma}(\sigma_{1})]
\int_{\sigma_{1}}^{\xi} d\xi' (\xi - \xi')
p_{2}(\xi-\xi')\widetilde{\gamma}(\xi') \; .
\label{eq:84}
\end{eqnarray}
For the kernel ${\cal K}^{(1)}(\xi,\xi')$ corresponding to
$\gamma^{(1)}(\xi',\xi'')$, we get from Eq.\ (\ref{eq:24})
\begin{equation}
{\cal K}^{(1)}(\xi,\xi') = {\cal K}^{(<)}(\xi,\xi')
\Theta(\sigma_{1}-\xi) + {\cal K}^{(>)}(\xi,\xi')
\Theta(\xi - \sigma_{1}) \;  ,
\label{eq:85}
\end{equation}
with
\begin{eqnarray}
{\cal K}^{(<)}(\xi,\xi') &=& p_{1}(\xi') - p_{1}(0) \nonumber \\ &+&
[1/\widetilde{\gamma}(\xi')] \left[ (\xi - \xi') p_{2}(\xi - \xi')
\widetilde{\gamma}(\xi) + \int_{\xi'}^{\xi} d\xi'' (\xi'' -\xi')
p_{3}(\xi''-\xi') \widetilde{\gamma}(\xi'') \right]
\label{eq:77}
\end{eqnarray}
and
\begin{eqnarray}
{\cal K}^{(>)}(\xi,\xi') &=& \left\{\rule{0mm}{6mm} [(\sigma_{1} -
\xi') p_{2}(\sigma_{1} - \xi') + p_{1}(\sigma_{1}-\xi') - p_{1}(\xi-\xi')]
[\widetilde{\gamma}(\sigma_{1})/\widetilde{\gamma}(\xi')] \right.  \nonumber \\
&+& \left.p_{1}(\xi') - p_{1}(0) + \int_{\xi'}^{\sigma_{1}} d\xi'' (\xi''-\xi')
p_{3}(\xi''-\xi') [\widetilde{\gamma}(\xi'')/\widetilde{\gamma}(\xi')]
\rule{0mm}{6mm}\right\} \Theta(\sigma_{1} - \xi') \nonumber \\ &+& \left\{
\rule{0mm}{6mm} [p_{1}(0) - p_{1}(\xi-\xi')] [\widetilde{\gamma}(\xi')/
\widetilde{\gamma}(\sigma_{1})] - (\xi' - \sigma_{1}) p_{2}(\xi' - \sigma_{1})
\right.  \nonumber \\ &+& \left.  p_{1}(\xi') - p_{1}(\xi' - \sigma_{1}) -
\int_{\sigma_{1}}^{\xi'} d\xi'' (\xi'-\xi'') p_{3}(\xi'-\xi'')
[\widetilde{\gamma}(\xi'')/ \widetilde{\gamma}(\sigma_{1})]\rule{0mm}{6mm}
\right\} \Theta(\xi' - \sigma_{1}) \; . \nonumber \\
\label{eq:86}
\end{eqnarray}
In the limits $\sigma_{1} \rightarrow \sigma$ and $\sigma_{1} \rightarrow 0$,
we obtain from Eqs.\ (\ref{eq:84}) and (\ref{eq:85}) the coefficient function
and kernel for monotonically rising profile and falling profile, respectively.

Considering specifically a piecewise {\em linear} profile with maximum $E_{\rm
c}^{0}$ at $S_{1}$, we write
\begin{equation}
E_{\rm c}(\xi) = E_{\rm c}^{0} \, h(l\xi) =  E_{\rm c}^{0} \, h(x)\; ,
\label{eq:86a}
\end{equation}
where
\begin{equation}
h(x) = \left\{ \begin{array}{ll} x/S_{1} & ; \; \; x \leq S_{1} \; , \\
(S-x)/(S-S_{1}) & ; \; \; x > S_{1} \; ,
\end{array}
\right.
\label{eq:86b}
\end{equation}
so that
\begin{equation}
\widetilde{\gamma}(\xi) = e^{-c_{0} h(l\xi)} \; ,
\label{eq:86c}
\end{equation}
with $c_{0} = \beta E_{\rm c}^{0}$.

\begin{figure}
\epsfysize=14.0cm
\epsfbox[66 206 503 623]{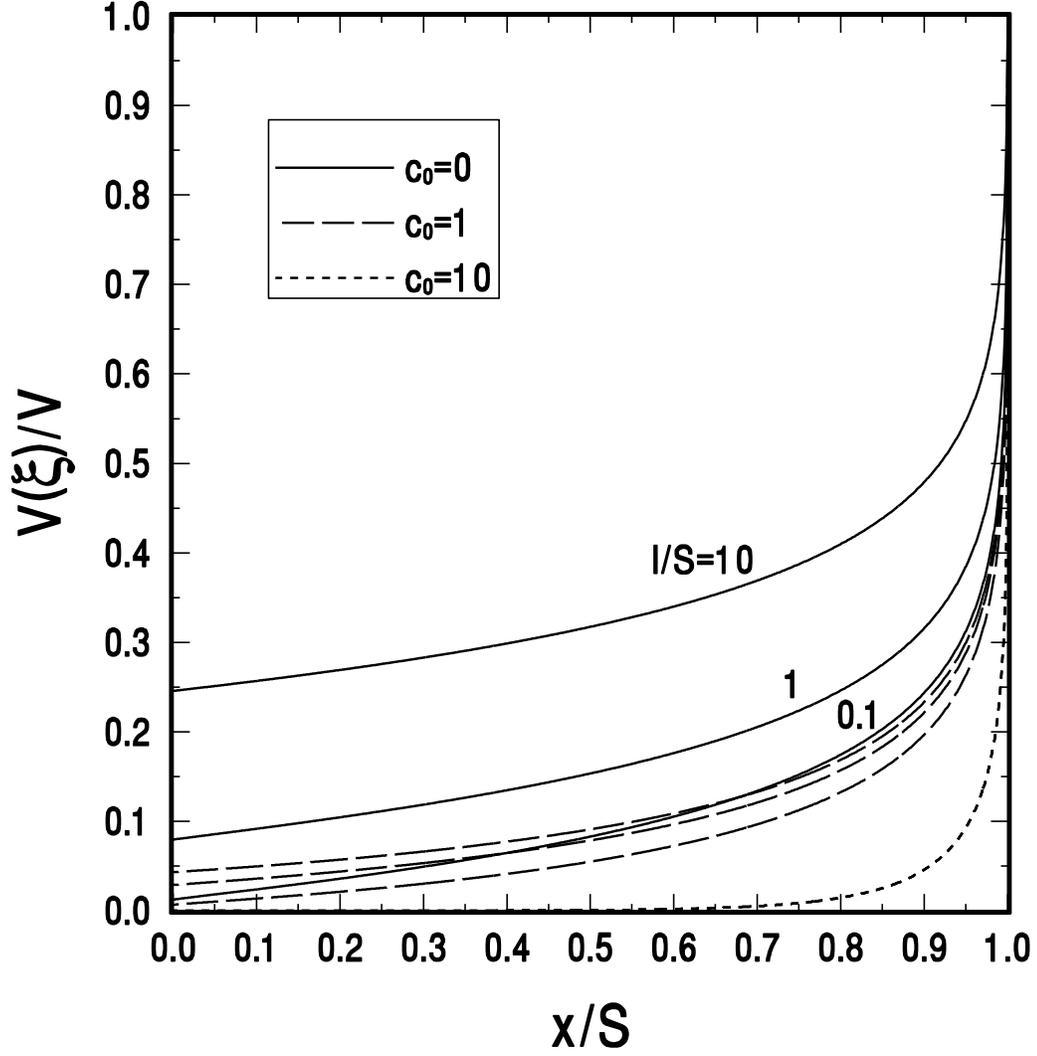}
\vspace*{2.0cm}
\caption{The normalized internal voltage difference $V(\xi)/V$ for
three-dimensional transport across a linearly rising band edge profile, plotted
as a function of the reduced coordinate $x/S = l\xi/S$ for various values of
$l/S$ and of the parameter $c_{0}$ ($c_{0} = 0$:\ flat profile [cf.\ Fig.\
\ref{fig:5}]; $\beta eV = eV/k_{\rm B}T = 10$).  Note that for $c_{0} = 10$,
the three curves corresponding to $l/S = 0.1$, $1$, and $10$ coincide within
the accuracy of the drawing.}
\label{fig:7}
\end{figure}

\begin{figure}
\epsfysize=14.0cm
\epsfbox[66 206 503 623]{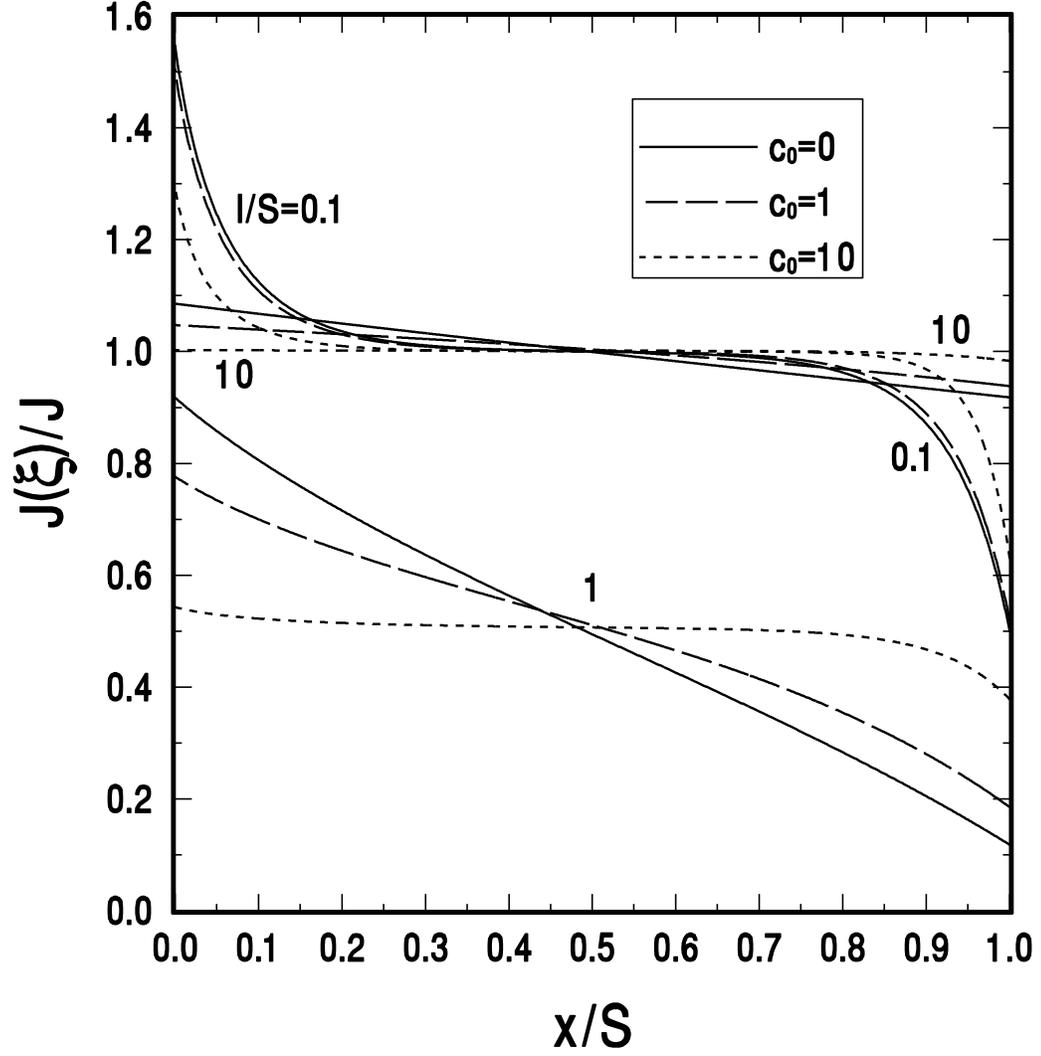}
\vspace*{2.0cm}
\caption{The normalized thermo-ballistic current $J(\xi)/J$ for
three-dim\-ensional transport across a linearly rising band edge profile,
plotted as a function of the reduced coordinate $x/S = l\xi/S$ for various
values of $l/S$ and of the parameter $c_{0}$ ($c_{0} = 0$:\ flat profile; cf.\
Fig.\ \ref{fig:6}).  For better distinguishability, the curves corresponding to
$l/S = 1$ have been shifted downwards by 0.5 units.}
\label{fig:8}
\end{figure}

In the limit $\sigma_{1} \rightarrow \sigma$, i.e., for a profile {\em rising
monotonically} over the length of the sample, the coefficient function and
kernel entering Eq.\ (\ref{eq:26a}) are given by the function $f^{(<)}$ of
Eq.~(\ref{eq:76}) and by the kernel ${\cal K}^{(<)}$ of Eq.\ (\ref{eq:77}),
respectively.  For this case, by numerically solving Eq.~(\ref{eq:26a}) for
three-dimensional transport, we have obtained the curves for the normalized
voltage difference $V(\xi)/V$ [i.e., for the quasi-Fermi level $E_{\rm
F}(\xi)$] shown in Fig.\ \ref{fig:7}, and for the normalized thermo-ballistic
current $J(\xi)/J$ shown in Fig.\ \ref{fig:8}.  The flat-profile results
($c_{0} = 0$) for three-dimensional transport (cf.\ Figs.\ \ref{fig:5} and
\ref{fig:6}) are included for comparison.

Two features of the curves shown in Fig.\ \ref{fig:7} are notable.  (i) With
increasing value of the parameter $c_{0}$, the quasi-Fermi level $E_{\rm
F}(\xi)$ becomes less sensitive to a change in the ratio $l/S$; for $c_{0} =
10$, the curves for the different values of $l/S$ virtually coincide.  (ii)
With increasing value of $c_{0}$, the curves tend to stay closer to the
abscissa, and the sharp rise in the curves is shifted progressively to the
vicinity of $x/S = 1$.  Both features can be explained by looking at expression
(\ref{eq:39}), in which the exponential term depending on the band edge profile
$E_{\rm c}(\xi)$ takes the form $\exp(-c_{0}[1-x/S])$ for the linearly rising
profile [cf.\ Eqs.\ (\ref{eq:86a}) and (\ref{eq:86b})].  For $c_{0} \gg 1$,
this term is sizeable only in the immediate vicinity of $x/S = 1$, where
$\chi(\xi) \approx \chi(S/l)$, independent of $S/l$. Thus, it emerges that in
the presence of a pronounced maximum in the profile, the entire internal
voltage change occurs in the vicinity of this maximum.

The localization of the voltage change near the maximum of the profile appears
to imply that the thermo-ballistic current $J(\xi)$ approaches the physical
current.  This follows from the results shown in Fig.\ \ref{fig:8} where the
normalized thermo-ballistic current$J(\xi)/J$ tends to approach unity when the
parameter $c_{0}$ is increased, regardless of the value chosen for $l/S$.

\begin{figure}
\epsfysize=14.0cm
\epsfbox[66 206 503 623]{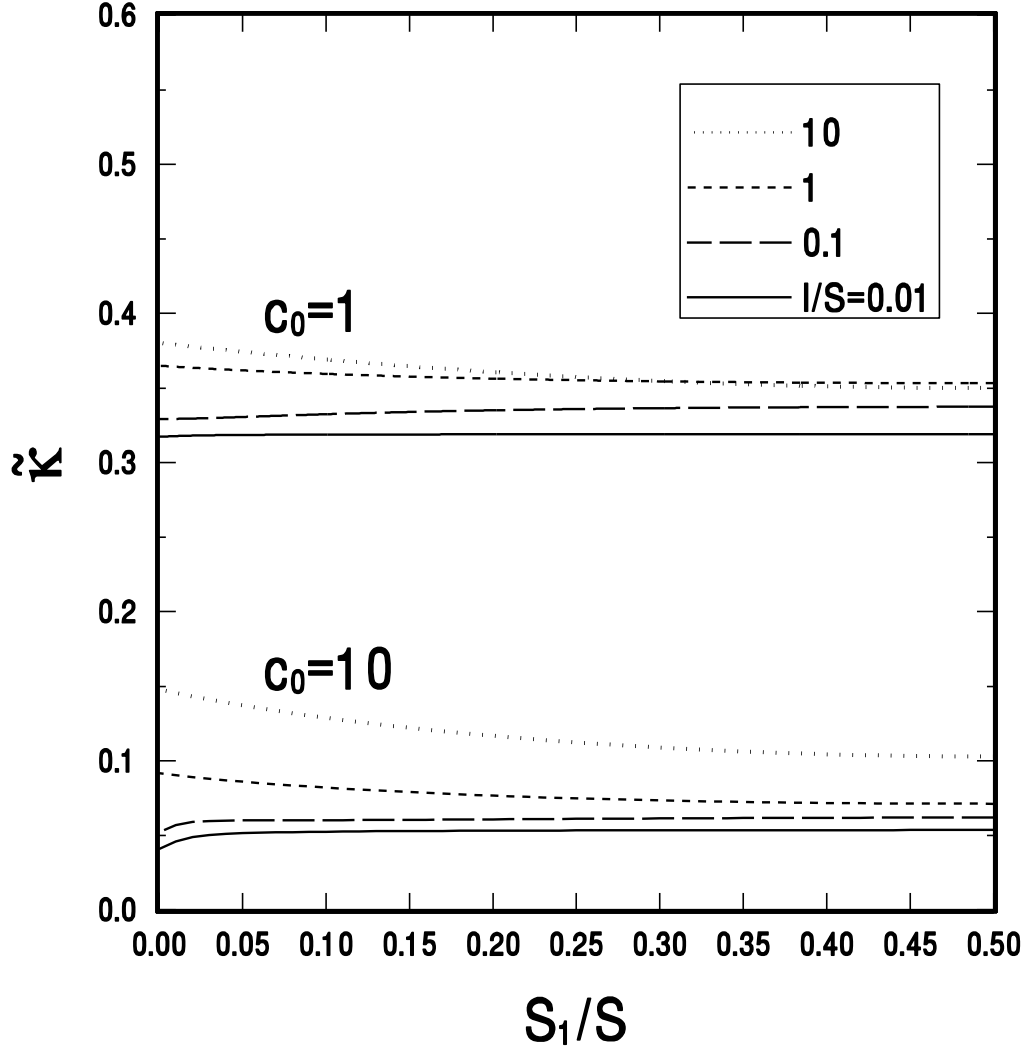}
\vspace*{2.0cm}
\caption{The quantity $\widetilde{\kappa}$ characterizing the reduced
resistance $\widetilde{\chi}$, calculated for one-dim\-ensional transport
across a band edge profile with a single maximum at $\xi = \sigma_{1} =
S_{1}/l$, and plotted as a function of $S_{1}/S$ for $c_{0} = 1$ and $10$ and
for various values of $l/S$.}
\label{fig:9}
\end{figure}

The effect of a variation of the position $S_{1}$ of the maximum is illustrated
in Fig.\ \ref{fig:9}. From the numerical solution of Eq.\ (\ref{eq:26a}) with
the coefficient function $f^{(1)}$ of Eq.\ (\ref{eq:83}) and the kernel ${\cal
K}^{(1)}$ of Eq.\ (\ref{eq:85}) for one-dimensional transport, we have obtained
the curves for the quantity $\widetilde{\kappa}$ [cf.\ Eqs.\ (\ref{eq:36bb})
and (\ref{eq:36c})] as a function of $S_{1}/S$, calculated for $c_{0}=1$ and
$10$ and for various values of $l/S$ (as the curves are symmetric about
$S_{1}/S = 0.5$, only the range $0 \leq S_{1}/S \leq 0.5$ is shown).
Generally, the dependence of $\widetilde{\kappa}$ on $S_{1}$ and on $l/S$ is
weak.  For $l/S \ll 1$ (diffusive limit), we have $\widetilde{\kappa}
\rightarrow (1/2) (\widetilde{S}/S) = (1/2)(1-e^{-c_{0}})/c_{0}$, independent
of $S_{1}$ [cf.\ Eqs.\ (\ref{eq:a13}) and (\ref{eq:a14})]; the latter is a
consequence of the fact that $\widetilde{S}$ is determined essentially by the
area below the profile, which is independent of the position of the maximum.
For $l/S \gg 1$ (ballistic limit), $\widetilde{\kappa}$ assumes its largest
values when the maximum of the profile lies at either end of the sample.
Remarkably, $\widetilde{\kappa}$ decreases with increasing value of the
parameter $c_{0}$, i.e., the reduced resistance $\widetilde{\chi}$ is {\em
lowered} as the height of the profile maximum is increased.  However, as
mentioned above, in the current-voltage characteristic this lowering of the
resistance is generally over-compensated by the barrier factor.

\begin{figure}
\epsfysize=14.0cm
\epsfbox[66 206 503 623]{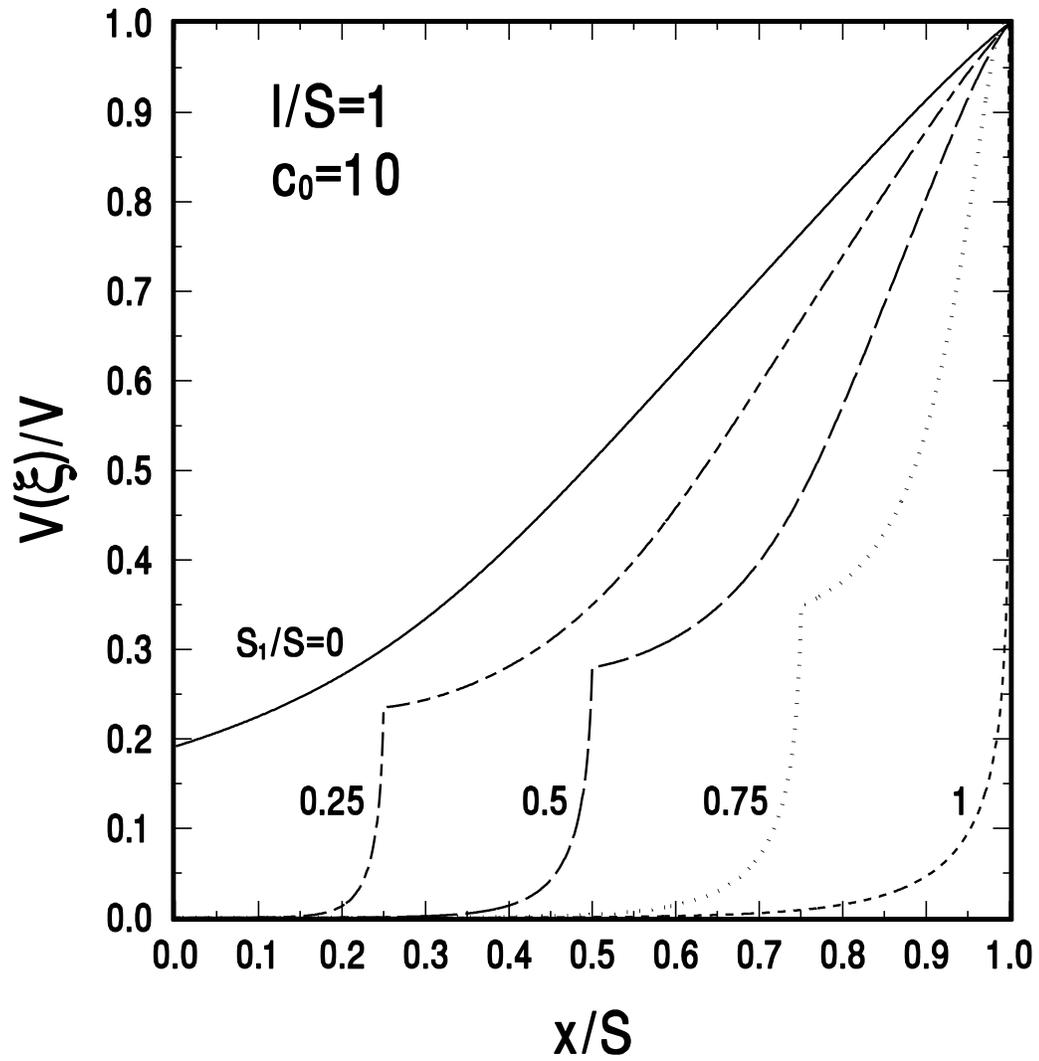}
\vspace*{2.0cm}
\caption{The normalized internal voltage difference $V(\xi)/V$, calculated for
one-dim\-ensional transport across a band edge profile with a single maximum at
$\xi = \sigma_{1} = S_{1}/l$, and plotted as a function of the reduced
coordinate $x/S = l\xi/S$ for $l/S=1$, $c_{0} = 10$, and various values of
$S_{1}/S$.}
\label{fig:10}
\end{figure}

In Fig.\ \ref{fig:10}, the normalized internal voltage difference $V(\xi)/V$ is
shown as a function of the reduced coordinate $x/S = l\xi/S$ for $l/S = 1$,
$c_{0} = 10$, and various values of $S_{1}/S$.  The value chosen for $c_{0}$
corresponds to a high profile maximum.  One observes an abrupt change of the
voltage difference in the vicinity of the maximum (cf.\ Fig.\ \ref{fig:7}).
The contact resistance seems to have been shifted into the sample unto whatever
the position of the maximum is.

\section{Comparative discussion}

In order to put the present description of carrier transport into proper
perspective, we compare it to previous studies attempting to unify ballistic
and diffusive transport.

An early synthesis of the ballistic and diffusive mechanisms for transport
across a single potential barrier has been put forward in Ref.\ \cite{cro66}
(cf.\ also Ref.\ \cite{sze81}).  The barrier maximum is enclosed in a narrow
ballistic interval, and the transport in the remainder of the sample is
regarded as diffusive.  For this special shape of profile, this {\em ad hoc}
model gives results qualitatively equal to ours.  It was first applied to
transport across metal-semiconductor contacts, and later (cf., e.g., Ref.\
\cite{eva91}) to transport across grain boundaries in polycrystalline
semiconducting materials.

In Ref.\ \cite{dej94}, the transition from the ballistic to the diffusive
transport regime in high-mobility wires has been studied through a transmission
approach based on the semiclassical limit of the Landauer transport formula
\cite{lan87}.  The zero-bias conductance has been obtained for homogeneous
samples (i.e., for flat potential profile) at zero temperature for
two-dimensional and three-dimensional transport.  The transmission problem has
been solved in analogy to the treatment of light scattering in a diffuse medium
\cite{mor53}.  Isotropic impurity scattering of the electrons has been assumed,
and the average transmission probability has been calculated within a two-step
formalism involving the solution of a Fredholm-type integral equation (Milne's
equation) and of a differential equation for the relevant transmission
probabilities.  The approach of Ref.\ \cite{dej94} thus differs from the
present one in that it is restricted to the case of flat profile at zero
temperature, and is based on a different physical picture.  As a consequence,
the results of Ref.\ \cite{dej94} generally deviate from those in the present
work; the conductance turns out to be the same for one-dimensional transport,
but is different for three-dimensional transport (cf.\ Fig.\ \ref{fig:4}).  An
interpolation formula proposed in Ref.\ \cite{dej94} has been used in Ref.\
\cite{pri98} to calculate the conductance across a single maximum in the band
edge profile (arising from a grain boundary in a polycrystalline
semiconductor).

In a previous paper by the present authors on the unification of
(one-dimensional) ballistic and diffusive carrier transport \cite{lip01}, a
closed-form expression for the reduced resistance was obtained, using ideas
similar to those which have led to the concept of the thermo-ballistic current
in the present paper.  In order to achieve this, it was assumed that the
current analogous to the thermo-ballistic current is conserved at the
equilibration points.  For arbitrary band edge profile, the reduced resistance
is then obtained in terms of explicit integrals over the profile, so that the
interplay of the effects of the profile and the mean free path can be studied
in a transparent manner.  The method of Ref.\ \cite{lip01} has proven useful in
the analysis of Hall mobility data for polycrystalline silicon
\cite{lip01a,wei02}; for a fully quantitative analysis, however, it should be
replaced with the present method.

The main feature of the present description is the partitioning of the physical
current into the thermo-ballistic current and a background current.  This
allows us to take into account the coupling to the thermal reservoirs, invoking
only global rather than local current conservation (the use of local
conservation of the thermo-ballistic current would actually lead to
inconsistent results).  In Refs.\ \cite{cro66,sze81,eva91,lip01}, local current
conservation was explicitly assumed, but it also underlies implicitly the
approach of Ref.\ \cite{dej94}.

In the calculation of the thermo-ballistic current via the solution of the
integral equation for the resistance function, as it is done in the present
work, the contact resistance appears in a natural way as an integral part of
the total resistance.  The resistance function, as well as the quasi-Fermi
level and the thermo-ballistic current determined by it, are auxiliary
functions which depend on the choice of reference contact.  This is a
well-known feature of the transmission approach to the conductance problem
\cite{dat95}. In the present approach, the physical results are independent of
the above-mentioned choice as a consequence of an appropriate symmetrization.

\section{Summary and outlook}

We have developed a unified semiclassical description of ballistic and
diffusive carrier transport in parallel-plane semiconductor structures.  A
novel theoretical formulation has been achieved by introducing the concept of a
thermo-ballistic current, which combines in a consistent manner the ballistic
motion of carriers in the band edge potential and their thermalization at
randomly distributed equilibration points.  Global conservation of the
thermo-ballistic current has been established by identifying its average over
the sample length with the physical current, for which the current-voltage
characteristic has been derived in terms of a reduced resistance.  The
formalism is applicable to the treatment of transport across arbitrarily shaped
band edge profiles.

Calculations of the reduced resistance have been carried out for simple band
edge profiles.  It has been found that generally the transport mechanism
changes smoothly from ballistic to diffusive as the value of the ratio of mean
free path and sample length changes from large to small.  The introduction of a
pronounced profile structure appears to weaken the dependence on this ratio.
The dimensionality of the electron motion turns out to have fairly small
effect, giving rise to a value of no more than 3/2 for the ratio of the reduced
resistances for three-dimensional and one-dimensional transport.  By using, in
one-dimensional calculations, the effective mean free path of the
three-dimensional case, one may closely approximate the results of
three-dimensional calculations.  Analyses of the local quasi-Fermi level and of
the thermo-ballistic current have provided detailed insight into the interplay
of the ballistic and diffusive transport mechanisms.

The description may be extended in several ways.  So far, we have regarded the
band edge profile as given, so that the integral equation for the resistance
function is linear; this is justified for weak bias.  In the general case, the
equation is nonlinear and must be solved self-consistently together with the
Poisson equation for the band edge profile.  Tunneling effects can be
incorporated in the formulation by changing the classical expression for the
transmission probabilities to the corresponding WKB expressions.
Inhomogeneously doped samples like, e.g., p-n junctions, can be treated by
introducing a position-dependent mean free path.  This involves a simple
generalization of the expressions for the probabilities of occurrence of the
ballistic intervals.  Another possible extension is to include the effect of
illumination.  To this end, the formalism is immediately adapted to the
transport of holes.  A mechanism for the absorption of light must be
introduced, which would eventually lead to the determination of the
photoconductance.  Implementing extensions of this kind in the unified
description of carrier transport should enable one to apply it in the analysis
of experiment data, e.g., for polycrystalline semiconducting materials where
carrier transport is governed by sequences of potential barriers arising from
carrier trapping at grain boundaries.

An extension of the present approach of great topical interest would be to
include the electron spin as an additional degree of freedom.  This would allow
us to treat spin-polarized transport and, thereby, to make contact with the
newly developing field of semiconductor spintronics \cite{aws02,joh02,sch02}.
There, the drift-diffusion approach has commonly been used \cite{ras02a,yuf02a}
to describe the transport of spin-polarized electrons.  Only very recently, a
synthesis of diffusive and ballistic spin transport has been attempted
\cite{kra03}.

In conclusion, we point out that the applicability of our approach is limited
mainly by two features.  First, the assumption of a one-dimensional geometry
leaves as principal fields of application the transport in parallel-plane
heterostructures (including two-dimensional wires) as well as in
polycrystalline semiconducting materials with a virtually one-dimensional
grain-boundary structure. Second, quantum coherence effects are beyond the
scope of the present semiclassical approach.  Therefore, limits are set to the
treatment of transport in nanostructures at low temperatures, where the phase
coherence lengths play a dominant role.

\begin{acknowledgments}

We are grateful to A.\ Ecker and B.\ Bohne for their assistance in the
numerical calculations and in the preparation of the figures.
\end{acknowledgments}

\appendix*

\section{Ballistic and diffusive limits}

For given band edge profile, we consider two limiting situations for
the mean free path $l$, the ballistic limit and the diffusive limit. To this
end, we rewrite Eqs.\ (\ref{eq:23}), (\ref{eq:24}), and (\ref{eq:26a}) in a
form exhibiting $l$ explicitly:
\begin{equation}
f(x;l) = \frac{x}{l} \, p_{1}(x/l) + \int_{0}^{x}  \frac{dx'}{l}\,
\frac{x-x'}{l} \, p_{2}((x-x')/l) \, \gamma (x',x) \; ,
\label{eq:a1}
\end{equation}
\begin{eqnarray}
{\cal K}(x,x';l) = - \frac{x'}{l} \, p_{2}(x'/l) &+& \frac{x-x'}{l} \,
p_{2}((x-x')/l) \, \gamma(x,x') \nonumber \\[0.3cm] &+& \int_{0}^{x}
\frac{dx''}{l}\, \frac{x''-x'}{l} \, p_{3}(|x'-x''|/l) \,
\gamma(x'',x') \; , \label{eq:a2}
\end{eqnarray}
\begin{equation}
\frac{x}{l} - f(x;l)\, \chi(x;l) + \int_{0}^{x} \frac{dx'}{l}
\, {\cal K} (x,x';l) \, \chi(x';l) = 0 \; .
\label{eq:a3}
\end{equation}

(i) In the {\em ballistic limit} $l/S \gg 1$, we write
\begin{equation}
\chi(x;l) \approx a(x) + b(x) \frac{S}{l} \; ,
\label{eq:a4}
\end{equation}
and find from Eqs.\ (\ref{eq:a1})-(\ref{eq:a3}), to second order in $S/l$,

\begin{eqnarray}
\frac{S}{l} &-& \left\{ \frac{S}{l} + p_{2}(0) \left( \frac{S}{l}\right)^{2}
\left[ -1 + \int_{0}^{S} \frac{dx}{S} \, (1-x/S) \, \gamma(x,S) \right]
\right\} \left[ a(S) + b(S)\frac{S}{l} \right] \nonumber \\[0.3cm] &\ &
\hspace{3.3cm} + \; p_{2}(0) \left( \frac{S}{l}\right)^{2}
 \int_{0}^{S} \frac{dx}{S} \, [-x/S + (1 - x/S)\,\gamma(S,x)]\, a(x) = 0
\; . \nonumber \\ \label{eq:a5} \end{eqnarray}
Collecting terms of order $S/l$, we obtain
\begin{equation}
a(S) = 1 \; ,
\label{eq:a6}
\end{equation}
while the terms  of order $(S/l)^{2}$ yield
\begin{equation}
b(S) = p_{2}(0) \left\{ \frac{1}{2} + \int_{0}^{S} \frac{dx}{S} \,
(1-x/S) \, [ \gamma(S,x) - \gamma(x,S)]\right\} \; .
\label{eq:a7}
\end{equation}
As a result, we arrive at
\begin{equation}
\widetilde{\chi}(l,[E_{\rm c}]) \approx 1 + p_{2}(0)\frac{S}{l} \left\{
\frac{1}{2} + \int_{0}^{S} \frac{dx}{S} \, (1-x/S) \, e^{-\beta
E_{\rm c}^{\rm m}(x,S)} \, [ e^{\beta E_{\rm c}^{\rm m}(0,x)} - e^{\beta E_{\rm
c}^{\rm m}(0,S)}] \right\} \; .
\label{eq:40}
\end{equation}
For $l/S \rightarrow \infty$, we have $\widetilde{\chi}(l,[E_{\rm c}]) = 1$, and
Eq.\ (\ref{eq:36}) goes over into the current-voltage characteristic in the
purely ballistic regime,
\begin{equation}
j \approx -e v_{\rm e} N_{\rm c} \, e^{-\beta E_{\rm p}} \, (1 - e^{- \beta
eV}) \; ;
\label{eq:41}
\end{equation}
in this regime, the current (or the conductance) is independent of the band
edge profile $E_{\rm c}$ (except for the height of its absolute maximum,
$E_{\rm c}^{\rm m}(0,S)$, in the interval $[0,S]$), and of the dimensionality
of the electron motion.

(ii) In the {\em diffusive limit} $l/S \ll 1$, we write
\begin{equation}
\chi(x;l) = \frac{B(x)}{l} \; ,
\label{eq:a8}
\end{equation}
where $B(x)$ is independent of $l$ to leading order. We then find from Eqs.
(\ref{eq:a1})-(\ref{eq:a3})
\begin{eqnarray}
x - 2 p_{0}(0)\, \gamma(x,x)\, B(x) &+& \int_{0}^{x} \frac{dx'}{l} \left[
\int_{0}^{x'} \frac{dx''}{l} \,  \frac{(x''-x')^{2}}{l} \,p_{3}((x'-x'')/l)) \;
\frac{\partial \gamma(x'',x')}{\partial x'} \right.
\nonumber \\[0.3cm]  &+& \left. \int_{x'}^{x} \frac{dx''}{l} \,
\frac{(x''-x')^{2}}{l} \,p_{3}((x''-x')/l)) \; \frac{\partial
\gamma(x'',x')}{\partial x'} \right] B(x')  = 0 \; . \nonumber \\
\label{eq:a9}
\end{eqnarray}
Closer inspection shows that for given $x'$, depending on whether $E_{\rm
c}(x') > E_{\rm c}(x'')$ or $E_{\rm c}(x') < E_{\rm c}(x'')$ when $x' \approx
x''$, either only the first or only the second of the two integrals over $x''$
contributes, either one yielding
\begin{equation}
2 p_{0}(0) \int_{0}^{x} dx' \, e^{\beta E_{\rm c}^{\rm m}(0,x')} \,
\frac{d}{dx'} e^{-\beta E_{\rm c}(x')} \; .
\label{eq:a10}
\end{equation}
We then have
\begin{equation}
x - 2 p_{0}(0) \left[ e^{\beta E_{\rm c}^{\rm m}(0,x)}  e^{-\beta E_{\rm
c}(x)} B(x) - \int_{0}^{x} dx' \,  e^{\beta E_{\rm c}^{\rm m}(0,x')} \, \left(
\frac{d}{dx'}  e^{-\beta E_{\rm c}(x')} \right) B(x') \right] = 0 \; .
\label{eq:a11}
\end{equation}
Taking the derivative with respect to $x$, we find
\begin{equation}
1 - 2 p_{0}(0)e^{-\beta E_{\rm c}(x)}\frac{d}{dx} [e^{\beta E_{\rm
c}^{\rm m}(0,x)} B(x)] = 0 \; ,
\label{eq:a12}
\end{equation}
so that
\begin{equation}
\chi(S;l) = \widetilde{\chi}(l,[E_{\rm c}]) \approx \frac{B(S)}{l} =
\frac{\widetilde{S}([E_{\rm c}])}{2p_{0}(0)l} \; , \label{eq:a13}
\end{equation}
where we have introduced the effective sample length
\begin{equation}
\widetilde{S}([E_{\rm c}]) = \int_{0}^{S} dx \, e^{-\beta [E_{\rm c}^{\rm
m}(0,S) - E_{\rm c}(x)]} \; .
\label{eq:a14}
\end{equation}
Equation (\ref{eq:a13}) implies for Eq.\ (\ref{eq:36})
\begin{equation}
j \approx -e v_{\rm e} N_{\rm c} \, e^{-\beta E_{\rm p}} \, \frac{2 \langle l
\rangle}{\widetilde{S}([E_{\rm c}])} \, (1 - e^{- \beta eV}) \; .
\label{eq:44}
\end{equation}
Thus, the current-voltage characteristic is seen to depend, in an integral way
via $\widetilde{S}([E_{\rm c}])$, on the band edge profile $E_{\rm c}(x)$, and on
the dimensionality of the electron motion via the effective mean free path
$\langle l \rangle = p_{0}(0)l$ [cf.\ Eq.\ (\ref{eq:9})]. Expression
(\ref{eq:44}) is equal to the result obtained for the stationary electron
current by integrating the drift-diffusion equation \cite{ash76}
\begin{equation}
j = \mu n(x) \, \frac{d}{dx} E_{\rm F}(x) \; ,
\label{eq:45}
\end{equation}
using Eq.\ (\ref{eq:2}); here, $\mu = 2e v_{\rm e} \beta \langle l \rangle$ is
the electron mobility.

It appears from Eq.\ (\ref{eq:a13}) that the criterion for the diffusive limit,
$l/S \ll 1$, should be more rigorously taken as $l/\widetilde{S} \ll 1$.

{}

\newpage

\newpage

\newpage

\newpage

\newpage

\newpage

\newpage

\newpage

\newpage

\newpage

\end{document}